\documentclass[intlimits,twoside,a4paper]{article}
\usepackage{amsmath,amssymb}
\usepackage{graphicx}
\usepackage{wrapfig}
\usepackage{xcolor}
\usepackage{lscape}

\usepackage[T2A]{fontenc}
\usepackage[cp1251]{inputenc}

\usepackage[eqsecnum]{cmpj}

\issue{2011}{14}{3}{33003}

\doinumber{10.5488/CMP.14.33003}

\title[Primitive model electrolytes]%
{Primitive model electrolytes. A comparison of the HNC approximation for
the activity coefficient \\ with Monte Carlo data%
\thanks{Dedicated to Professor Yura V.~Kalyuzhnyi on the occasion of his
60$^\text{th}$ birthday.}}
\author[E.~Guti\'{e}rrez-Valladares et al.]{E.~Guti\'{e}rrez-Valladares\refaddr{label1,label2},
M.~Luk\v{s}i\v{c}\refaddr{label2},
B.~Mill\'{a}n-Malo\refaddr{label1}, B.~Hribar-Lee\refaddr{label2},
V.~Vlachy\refaddr{label2}}
\addresses{
\addr{label1} Centro de F\'{i}sica Aplicada y Tecnolog\'{i}a
Avanzada, Universidad Nacional Aut\'{o}noma de M\'{e}xico,\\ A.P.
1--1010, 76000~Quer\'{e}taro, M\'{e}xico
\addr{label2} University of Ljubljana, Faculty of Chemistry and
Chemical Technology,\\ A\v{s}ker\v{c}eva c. 5, SI--1000 Ljubljana,
Slovenia}

\authorcopyright{E.~Guti\'{e}rrez-Valladares,
M.~Luk\v{s}i\v{c}, B.~Mill\'{a}n-Malo, B.~Hribar-Lee,
V.~Vlachy, 2011}

\date{Received April 15, 2011, in final form June 16, 2011}

\begin{document}

\maketitle

\begin{abstract}
Accuracy of the mean activity coefficient expression
(Hansen-Vieillefosse-Belloni equation), valid within the
hypernetted chain (HNC) approximation, was tested in a wide
concentration range against new Monte Carlo (MC) data for $+1$:$-1$
and $+2$:$-2$ primitive model electrolytes. The expression has an
advantage that the excess chemical potential can be obtained
directly, without invoking the time consuming Gibbs-Duhem
calculation. We found the HNC results for the mean activity
coefficient to be in good agreement with the machine calculations
performed for the same model. In addition, the thermodynamic
consistency of the HNC approximation was tested. The mean activity
coefficients, calculated via the Gibbs-Duhem equation, seem to
follow the MC data slightly better than the
Hansen-Vieillefosse-Belloni expression. For completeness of the
calculation, the HNC excess internal energies and osmotic
coefficients are also presented. These results are compared with
the calculations based on other theories commonly used to describe
electrolyte solutions, such as the mean spherical approximation,
Pitzer's extension of the Debye-H\"uckel theory, and the
Debye-H\"uckel limiting law.
\pacs 02.30.Rz, 05.10.Ln, 05.20.Jj, 05.70.Ce, 82.45.Gj, 82.60.Lf
\end{abstract}

\section{Introduction}

Knowledge of the physico-chemical properties of electrolyte
solutions contributes toward better understanding of the processes
in biology, chemical sciences, and various technologies. It is of
no surprise that electrolyte solutions are, after a hundred years,
still extensively studied in search of theories providing
quantitative description of these systems. The advances in
modeling, together with the review of experimental data for
electrolyte solutions, are summarized in books and thematic papers
(see, for example,
references~\cite{harned-owen,robinson-stokes,barthel88,Bockris1998,loebe97,Lamm2003}).
Although it has been established that only the models that
explicitly include solvent are capable of providing a proper
microscopical description of electrolyte
solutions~\cite{fennell,holovko}, one can describe thermodynamic
properties of these systems on the McMillan-Mayer level of
approximation by a proper choice of ion size
parameters~\cite{simonin98,hribar05}. In this paper we will
restrict ourselves to the latter case.

Seminal efforts to understand the properties of electrolyte
solutions were made by Debye and H\"{u}ckel (DH)~\cite{debye27}.
The DH theory is based on a linearized version of the
Poisson-Boltzmann equation which yields important insights into
dilute electrolyte solutions. Unfortunately, this approach can
only be used to quantitatively describe very dilute solutions of
size symmetric $+1$:$-1$ electrolytes~\cite{valleau80}. The
deficiencies of the Debye and H\"{u}ckel theory were analyzed in
many contributions (see, e.g.,
references~\cite{harned-owen,barthel88,friedman62,Outhwaite1975,Fixman,abbas07}).
Various continuations were suggested to extend the range of
applicability of the theory, among these the one proposed by
Pitzer~\cite{pitzer77} is very useful for engineering purposes.
This approach is simple, fully analytical, and capable of
semi-quantitatively describing the osmotic coefficient behavior of
$+1$:$-1$ electrolyte up to the concentration close to
$1.0$~mol\,dm$^{-3}$~\cite{pitzer77}.

Theoretical extension, correcting for the deficiency of the Debye
and H\"{u}ckel theory, is provided by the modified
Poisson-Boltzmann
approach~\cite{Outhwaite1975,Martinez1990,Outhwaite1991,Outhwaite1993,Outhwaite2004}.
For example, the individual activity coefficients of pure
electrolytes were calculated by Molero et al. (cf.
reference~\cite{Molero1992}) and for ternary systems containing
neutral hard spheres recently by Outhwaite et
al.~\cite{Outhwaite2010}. The theory yields excellent agreement
with computer simulations.

Another approach to calculate the properties of solutions involve
a class of integral equation theories based on the
Ornstein-Zernike (OZ) integral
equation~\cite{hansen06,mcquarrie73} and approximate closures. The
so-called mean spherical approximation (MSA) was solved
analytically for the primitive model electrolytes, with the
thermodynamic quantities written in a closed
form~\cite{blum75,sanchezcastro89,ebeling83,waisman72,blum77,triolo76,triolo78,corti87}.
Through the energy route, one obtains the osmotic and activity
coefficients that are in good agreement with Monte Carlo computer
simulations for $+1$:$-1$ electrolytes~\cite{valleau80}. Due to its
analytical nature, the MSA theory provides useful insights and is
accordingly extensively used (see, e.g.,
references~\cite{cartailler92,vilarino01,fawcett97,vilarino97,vilarino96,vilarino99,vilarino97_jsc}).
Another approximate but very robust closure of the
Ornstein-Zernike theory was proposed by Kovalenko and Hirata
(KH)~\cite{Kovalenko1999,Schmeer2010}. By changing the radii of
the ions~\cite{triolo78,Simonin1996,Fawcett1996,Vincze2010} and/or
including the association between unlike
ions~\cite{Blum1995,Bernard1996} these methods provide good fits
to experimental data. An integral equation approximation that is
widely used in describing the thermodynamics of symmetric, as well
as highly asymmetric electrolytes, is the hypernetted-chain (HNC)
closure~\cite{hansen06,valleau80,Rasaiah1988,vlachy99}.

An advantage of the OZ hypernetted-chain approximation is that it
yields accurate structural description of electrolyte solutions in
terms of the pair-distribution functions, $g_{ij}(r)$. Once this
information is known, the standard statistical-mechanical
equations, connecting the pair-distribution functions with
thermodynamic properties, can be applied~\cite{hansen06}. The
reduced excess internal energy of the system is readily obtained
via the expression
\begin{equation}
\frac{\beta E^{\mathrm{ex}}_{\mathrm{HNC}}}{N} = \frac{\beta}{2}
\sum_{i} \sum_{j} \frac{\rho_i\rho_j}{\rho} \int
u_{ij}(r)g_{ij}(r) \mathrm{d}{\bf r}, \qquad (i,j = +,-)
\label{energy}
\end{equation}
where $\beta = 1/k_\mathrm{B}T$ ($k_\mathrm{B}$ being the
Boltzmann constant and $T$ the absolute temperature), $\rho_i$
($\rho_j$) are the number densities of the species $i$ ($j$),
$\rho$ is the total number density of the system, $u_{ij}(r)$
the interaction pair potential (see section~\ref{model}), and
$\mathrm{d}{\bf r} = 4\pi r^2 \mathrm{d}r$.

The osmotic coefficient, $\Phi_\mathrm{HNC}$\,, can be
conveniently calculated using the virial route~\cite{hansen06}.
For the primitive electrolyte, the pair potential contains a
discontinuity at the distance of closest approach of two ions (see
section~\ref{model}). The problem can be surmounted by the
introduction of the so-called background correlation (referred to
as the cavity distribution) function, $y_{ij}(r) = \exp [\beta
u_{ij}(r) ] g_{ij}(r)$, and the formal division of the $u_{ij}(r)$
to the Coulombic, $u_{ij}^{\mathrm{C}}(r)$, and the hard sphere
part. In this way, the equation gets the form applicable to the
primitive model electrolyte~\cite{hansen06}
\begin{equation}
\Phi_\mathrm{HNC} = 1 - \frac{\beta}{6} \sum_{i} \sum_{j}  \frac{\rho_i \rho_j}{\rho} \int
r \frac{\mathrm{d}u_{ij}^{\mathrm{C}}(r)}{\mathrm{d}r} g_{ij}(r) \mathrm{d}{\bf r}
+ \frac{2\pi}{3} \sum_{i} \sum_{j}
 \frac{\rho_i \rho_j}{\rho} a_{ij}^3 g_{ij}(a_{ij}), \quad (i,j=+,-).
\label{eq:osmoznikoeficient1}
\end{equation}
Here $a_{ij}$ denotes the distance of closest approach of two
spherical particles $i$ and $j$, and $u_{ij}^{\mathrm{C}}(r)
\propto r^{-1}$ for all $r$, is the usual Coulomb potential.

The mean activity coefficient, $\gamma _{\pm}$\,, can be obtained
by integration of the Gibbs-Duhem equation~\cite{robinson-stokes}
\begin{equation}
\ln \gamma _{\pm,\mathrm{GD}} = \Phi_{\mathrm{HNC}} -1 + \int_{c=0}^{c} (\Phi_{\mathrm{HNC}} -1) \mathrm{d} \ln c' ,
\label{gd}
\end{equation}
where $c$ denotes the molar concentration of the electrolyte
solution.

Though the numerical procedure looks straightforward, it may be
time consuming: (i) the osmotic coefficients need to be obtained
for the series of concentrations, and (ii) for electrolyte
concentrations lower than $c=0.001$~mol\,dm$^{-3}$, the integral
in equation~(\ref{gd}) may diverge due to the numerical problems.
This particular problem can be resolved by using the analytical
continuation, for example DHLL+B2, at very low electrolyte
concentrations~\cite{martinez90}. The mean activity coefficient
calculation {via} the Gibbs-Duhem equation becomes particularly
cumbersome and time consuming in ternary mixtures containing
additional solutes~\cite{vlachy10}. The expressions written above
are general and can be in principle, apart from the HNC theory,
applied to any other approximate closure.

An alternative way of calculating the activity coefficients, but
valid only within the HNC approximation, has been proposed by
Verlet and Levesque~\cite{verlet62}. In its current form the
formula was  written by Hansen and Vieillefosse~\cite{hansen76},
and was successfully applied by Belloni~\cite{belloni85} for
asymmetric electrolytes. The expression, here referred to as the
Hansen-Vieillefosse-Belloni (HVB) equation, reads
\begin{equation}
\ln \gamma_{i,\mathrm{HVB}} = - \sum_{j}\rho_j {\bf
{c}_{ij}^{(s)}}(0) + \frac{1}{2} \sum_{j}\rho_j \int \left \{
h_{ij}(r) \left [ h_{ij}(r) - c_{ij}(r) \right ] \right \}
\mathrm{d}{\bf r}, \qquad (j=+,-). \label{belloni}
\end{equation}
Here $\gamma _{i}$ is the individual activity coefficient of the
species $i$ ($+$ or $-$), $h(r)$ denotes the total, and $c(r)$ the
direct  correlation function, while ${\bf c^{(s)}}(0)$ denotes the
Fourier transform of the short-range part of the direct
correlation function at $k=0$. Further, $\rho_j$ is the number
density of species $j$.

To our best knowledge, the accuracy of the activity coefficient
expression given by HVB equation~(\ref{belloni}) has not been
thoroughly tested for the primitive model electrolytes. In a few
cases, where the Monte Carlo data for the same system were
available~\cite{jamnik95}, the expression was found to be in good
agreement with the ``exact'' machine calculations. Therefore, the
purpose of this work is  to systematically examine the validity of
HVB equation~(\ref{belloni}) for the size symmetric and asymmetric
$+1$:$-1$, and size symmetric $+2$:$-2$ electrolytes. Notice that the
formula given by equation~(\ref{belloni}) is only valid within the
HNC approximation and therefore is not  generally applicable. In
other words, different expressions for the mean activity
coefficients can be derived consistently with the closure
conditions used~\cite{Kjellander1989}.

Even so, the HVB equation is very useful because it avoids long
and cumbersome evaluation of the excess chemical potential of the
solute {via} the Gibbs-Duhem route (see, e.g.,
reference~\cite{vlachy10}). Notice that the mean activity
coefficient of bulk electrolyte is often an input information for
studying heterogeneous systems, electrical double-layer, or Donnan
equilibrium, i.e. wherever the bulk electrolyte is in equilibrium
with charged surfaces. The results obtained by means of HNC
approximation, coupled with the HVB equation, are in the present
study compared with the new Monte Carlo simulations, and with the
results of some other electrolyte theories used in describing
electrolyte solutions. In addition, the internal consistency of
the HNC theory is tested with respect to the two different routes
of the mean activity coefficient calculation.

The paper is organized as follows: following a brief introduction,
the model and methods are presented, followed by the results
and discussion section. Numerical results are presented in form
of figures and tables.  Conclusions are given at the end.
Appendices summarize all the relevant equations used in these
calculations: Debye-H\"{u}ckel theory (appendix~A), Pitzer's
approach (appendix~B), and the mean spherical approximation
study (appendix~C).

\section{Model and methods \label{model}}

The model most frequently used in describing the electrolyte
solutions is called the primitive model. In this description, an
ion is represented as a hard sphere carrying positive or negative
charge in its center, while the solution as a whole is treated as
a dielectric continuum with a relative permittivity
$\varepsilon_{\rm r}$ (McMillan-Mayer level of approximation) of
pure solvent at pressure ($p$) and temperature ($T$) of
observation~\cite{hansen06}. Despite its simplicity, the model is
capable of explaining many experimentally determinable properties
of real electrolytes~\cite{simonin98,abbas09}, as well as their
mixtures~\cite{hribar05}.

The interaction pair potential between two ions of valence
$z_i$ ($z_j$), separated by a distance $r$, contains the hard
sphere and Coulomb part, $u_{ij}^{\mathrm{C}}(r)$
\begin{equation}
\beta u_{ij}(r)= \left \{
\begin{array}{cc}
\infty , & r < a_{ij}\,, \\[6pt]
\beta u_{ij}^{\mathrm{C}}(r)=z_{i} z_{j}
\frac{\lambda_{\mathrm{B}}}{r}\,, & r \geqslant a_{ij}\,,
\end{array}
\right.
\label{coulomb}
\end{equation}
where  $a_{ij}=\frac{1}{2}(a_{i}+a_{j})$, $a_{i}$ ($a_{j}$) is
the diameter of species $i$ ($j$), and $\lambda_\mathrm{B}$ is
the Bjerrum length
\begin{equation*}
\lambda_{\mathrm{B}}=\frac{\beta
e_0^{2}}{4\pi\varepsilon_{0}\varepsilon_{\rm r}}\,.
\end{equation*}
Here $\beta=1/k_\mathrm{B}T$, where $k_\mathrm{B}$ is the
Boltzmann constant, $T$ is the absolute temperature,
$\varepsilon_{0}$ is the permittivity in vacuum and $e_0$ is the
elementary charge. For aqueous solutions at 25~$^{\circ}$C
studied here, the Bjerrum length assumes the value
$\lambda_{\mathrm{B}}=7.14$~\AA.  Three different models of
aqueous electrolyte solutions at 25~$^{\circ}$C were
considered: (a) $z_{+}= - z_{-} = 1$, $a_{+}=a_{-} = 4.25$~\AA;
(b) $z_{+}= - z_{-} = 1$, $a_{+}= 5.43$~\AA, $a_{-} =
3.62$~\AA; and (c) $z_{+}= - z_{-} = 2$, $a_{+}=a_{-} =
4.25$~\AA. These values are often used as representatives for
simple ions in aqueous solutions.

\subsection{Monte Carlo simulation}

The Monte Carlo calculations were performed in the canonical
and grand canonical (GCMC) ensemble, using the standard
Metropolis sampling algorithm~\cite{allen_tildesley}. Periodic
boundary conditions with the minimum image (MI) convention, and
Ewald summation (ES) method were used in the canonical ensemble
simulations to minimize the finite sample effects, while GCMC
was performed only in combination with the Ewald summation. The
calculations were carried out with equal number of cations and
anions, the total number of particles $N$ being between 200 and
1000. After an equilibration run of $(1-10) \cdot 10^{7}$
configurations, each production run consisted of $(1-10) \cdot
10^{7}$ attempted configurations. From four to ten independent runs were performed
for simulation in canonical ensemble for each
concentration studied. The results given in tables and figures
are the average values over these runs, after the equilibration
runs were discarded.

While the excess internal energy and osmotic coefficient were
obtained as simple canonical ensemble averages~\cite{allen_tildesley},
the Widom's test particle insertion method~\cite{widom63,svensson91} was
used in the canonical ensemble simulation to calculate the excess
chemical potential of the electrolyte in solution
\begin{equation}
\beta \mu_{\pm} ^{\mathrm{ex}} = \ln \gamma_{\pm} = - \frac{1}{2}
\ln \langle \exp [-\beta U_{\mathrm{pair}}] \rangle. \label{widom}
\end{equation}
In the expression above, $\langle\cdots\rangle$ denotes the
canonical ensemble average, while $U_{\mathrm{pair}}$ is the
interaction energy between the system and a non-perturbing test
pair of oppositely charged ions, inserted at random locations in
the system. During the canonical ensemble simulation, for every
$N$ (number of particles in the simulation box) configuration, a
hundred of (Widom's) insertions were attempted, to obtain the
canonical ensemble average requested by equation~(\ref{widom}).

The results for the mean chemical potential of the electrolyte
solution as obtained by Widom's method may, at higher
concentrations, depend on the number of particles in the
system~\cite{svensson1988}. For this reason, in many cases,
especially for more concentrated electrolytes, the canonical ensemble
simulations were supplemented by the grand canonical ensemble Monte Carlo method.
In the latter case, the excess chemical potential is obtained
directly, without invoking the insertion
method~\cite{valleau1980,jamnik1993}.

\subsection{Hypernetted chain approximation}

The hypernetted chain (HNC) approximation is based on the
Ornstein-Zernike (OZ) equation. For multi-component mixtures, the
OZ equation reads~\cite{hansen06}
\begin{equation}
h_{ij}(r) = c_{ij}(r) + \sum_{k} \rho_{k} \int h_{ik}(|{\bf r} -
{\bf r}^\prime|) c_{kj}({\bf r}^\prime) \mathrm{d}{\bf r}^\prime ,
\label{oz}
\end{equation}
where $h(r)$ and $c(r)$ are the total and direct correlation
functions, respectively, and the integral is of the convolution
type. A general closure relation between $h(r)$ and $c(r)$ for
the OZ equation is~\cite{Rasaiah1988}
\begin{equation}
\ln \left [ h_{ij}(r) + 1 \right ] = -\beta u_{ij}(r) + h_{ij}(r)
- c_{ij}(r) + \mathcal{B}_{ij}(r). \label{hnc}
\end{equation}
$\mathcal{B}_{ij}(r)$ is in literature known as the ``bridge
graph'' and cannot be written as a closed form function of the
distribution functions $h(r)$ and $c(r)$. $\mathcal{B}_{ij}(r)$
are all the graphs in the representation of the background
correlation function, which are neither series nor parallel
graphs. At least two field points are bridged, i.e. connected by
the Mayer bond. The HNC approximation assumes that these graphs
mutually cancel and sets $\mathcal{B}_{ij}(r)$ to zero. Due to the
long-range nature of the Coulomb interaction, the set of
equations~(\ref{oz}) and~(\ref{hnc}) has to be re-normalized
before it can be solved numerically~\cite{Rasaiah1988,ichiye88}.

The re-normalized form of the integral equation was solved by
direct iteration using the fast Fourier transform routine on a
linear grid with $2^{18}$ division points separated by the
distance of $\Delta r$=0.005~\AA. It should be noted that the
results strongly depend on the number of points and separation
interval $\Delta r$. The decisive criteria in our case was the
smallest zeroth (electroneutrality condition) and second moment
condition that should both in theory be equal to
zero~\cite{Rasaiah1970,Stillinger1968}.

Equations~(\ref{energy}) and~(\ref{eq:osmoznikoeficient1}) were
used to calculate the excess internal energy and the osmotic
coefficients, respectively, while the equation~(\ref{belloni}) was
used to calculate the activity coefficient. The mean activity
coefficient, $\gamma_{\pm}$\,, for charge symmetric electrolytes
studied here, is finally calculated from the expression
$\gamma_{\pm} ^2 = \gamma_{+} \gamma_{-}$\,. In order to check the
thermodynamic consistency of the HNC approximation, the
\textit{mean} activity coefficients were (for a limited number of
cases) calculated with the help of the Gibbs-Duhem
equation~(\ref{gd}), using the HNC osmotic coefficient
(equation~(\ref{eq:osmoznikoeficient1})) results.

\section{Results and discussion}\label{results}

\begin{figure}[!t]
\includegraphics[width=5.2cm, angle=-90]{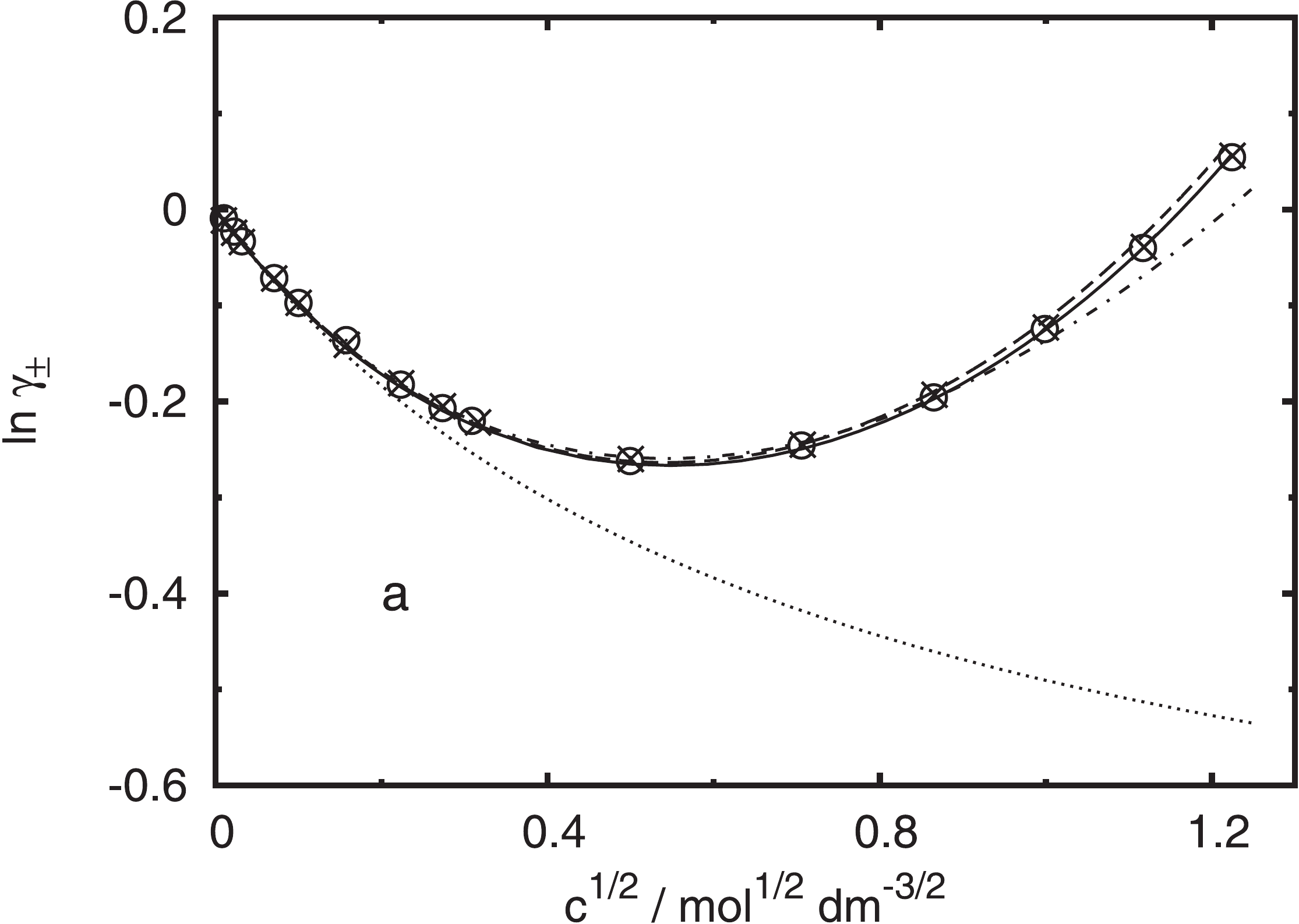}
\hfill
\includegraphics[width=5.2cm, angle=-90]{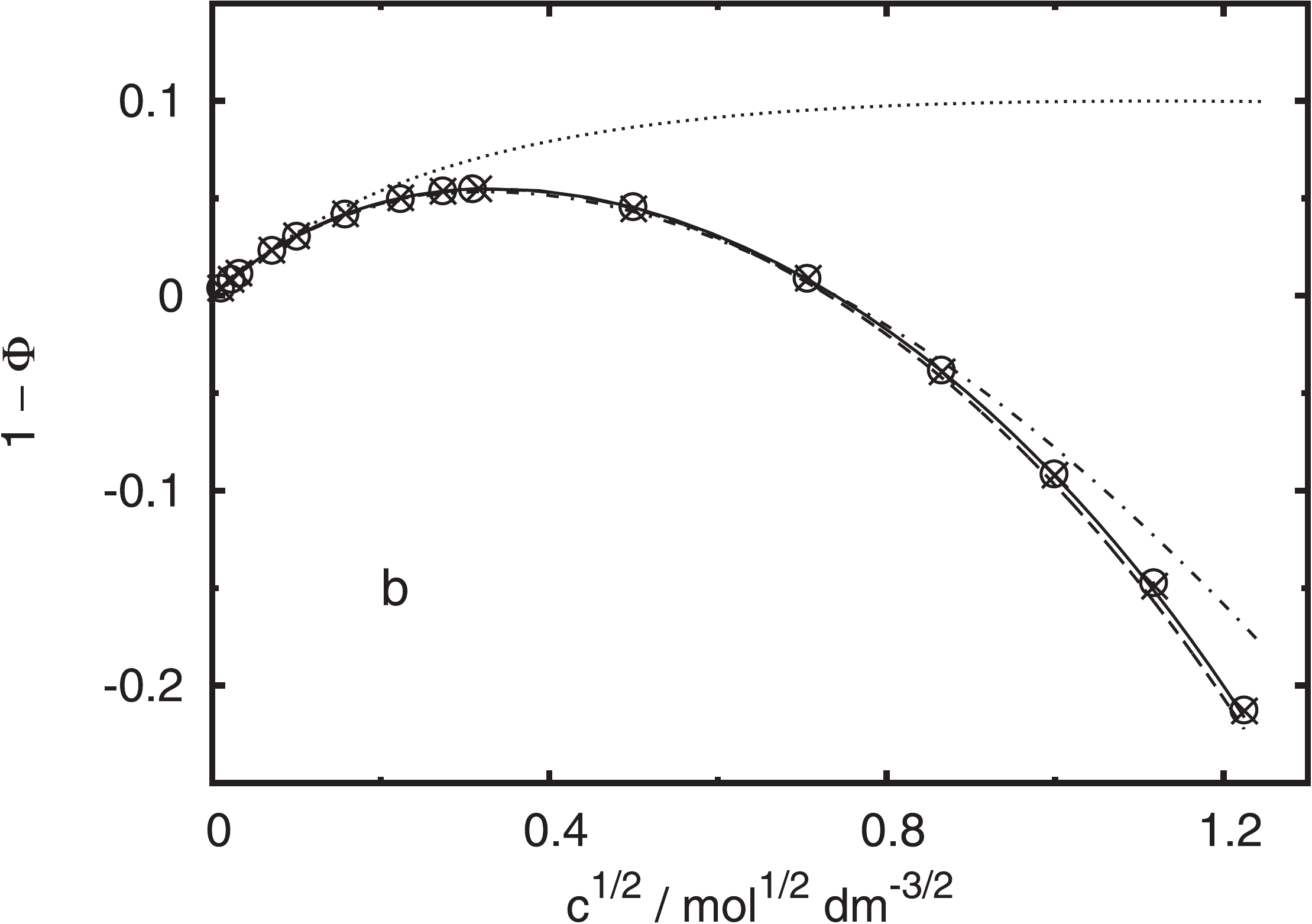}
\caption{$\ln \gamma_{\pm}$ (panel (a)) and $1-\Phi$ (panel (b))
as a function of $c^{1/2}$ for a $+1$:$-1$ primitive model
electrolyte, $a_+ = a_- = 4.25$~\AA, $\lambda_\mathrm{B} =
7.14$~\AA. Circles are grand canonical Monte Carlo data, crosses
represent the canonical ensemble Monte Carlo results (Ewald
summation). Theoretical values are given by lines: continuous
line~-- HNC, dashed line~-- MSA, dash-dotted line~-- Pitzer, and
dotted line~-- DH.} \label{fig:11sim}
\end{figure}

\begin{figure}[!h]
\includegraphics[width=5.2cm, angle=-90]{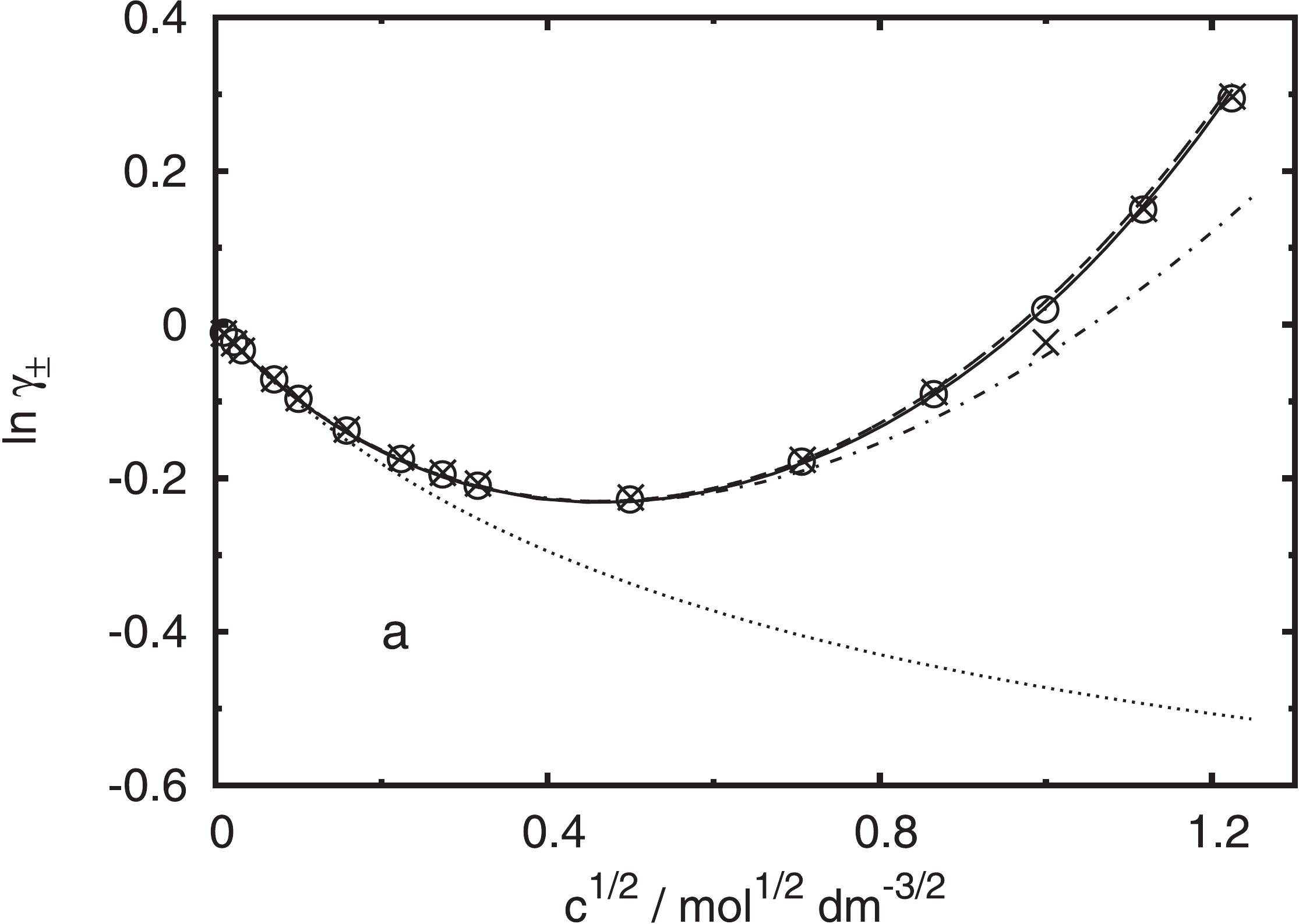}
\hfill
\includegraphics[width=5.2cm, angle=-90]{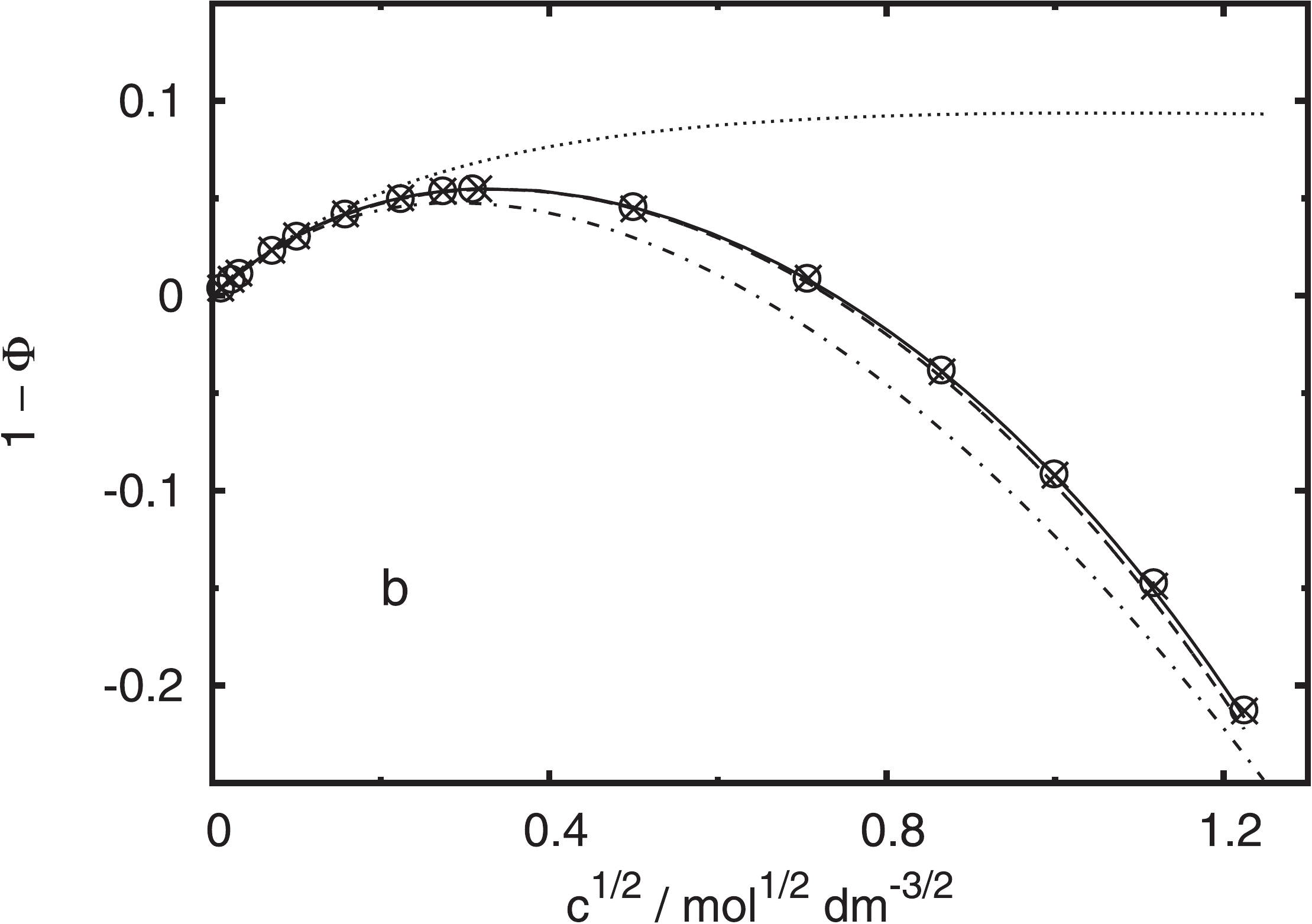}
\caption{Same as in figure~\ref{fig:11sim}, but for $a_+ =
5.43$~\AA\ and $a_- = 3.62$~\AA.} \label{fig:11asim}
\end{figure}

\begin{figure}[!h]
\includegraphics[width=5.2cm, angle=-90]{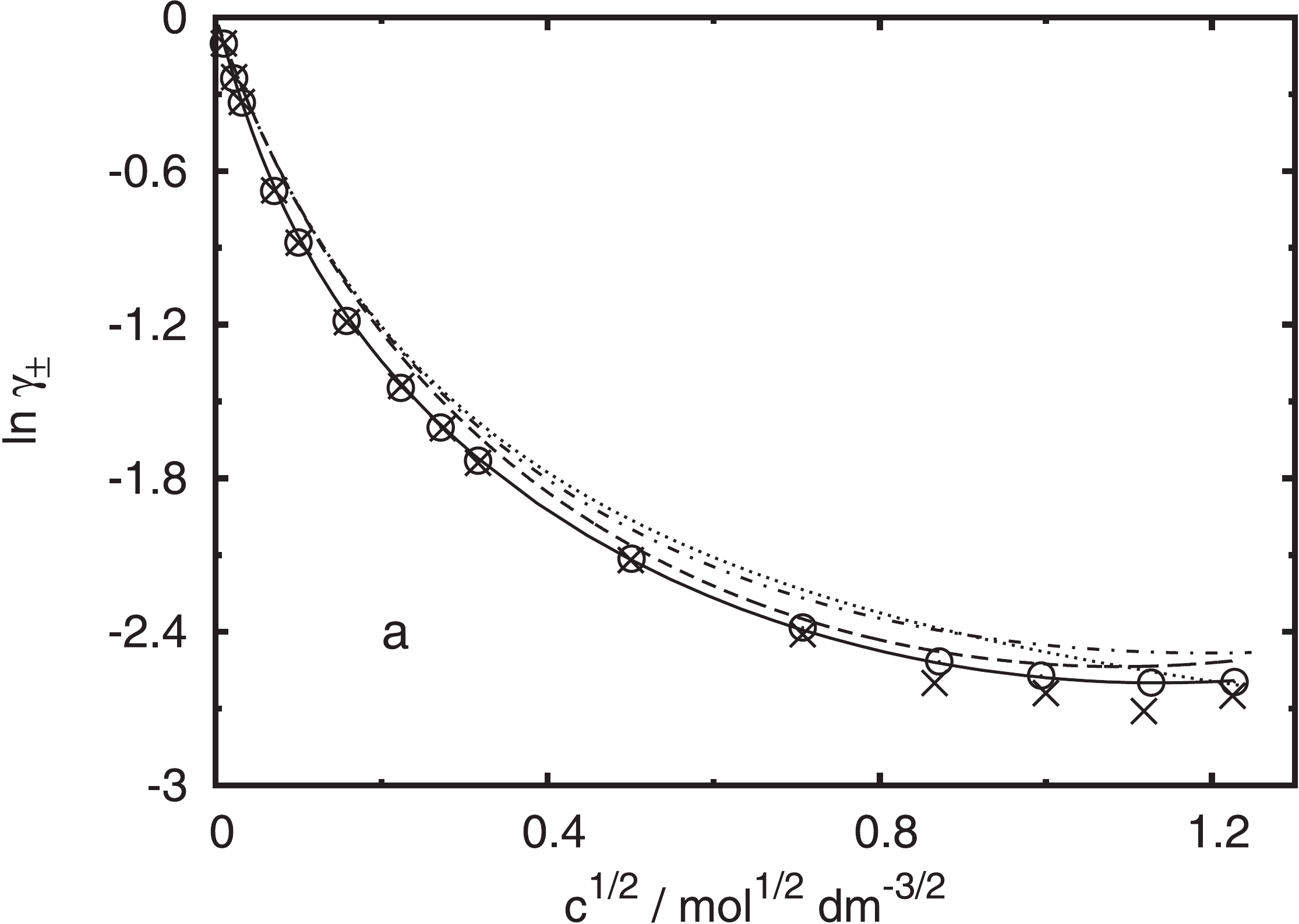}
\hfill
\includegraphics[width=5.2cm, angle=-90]{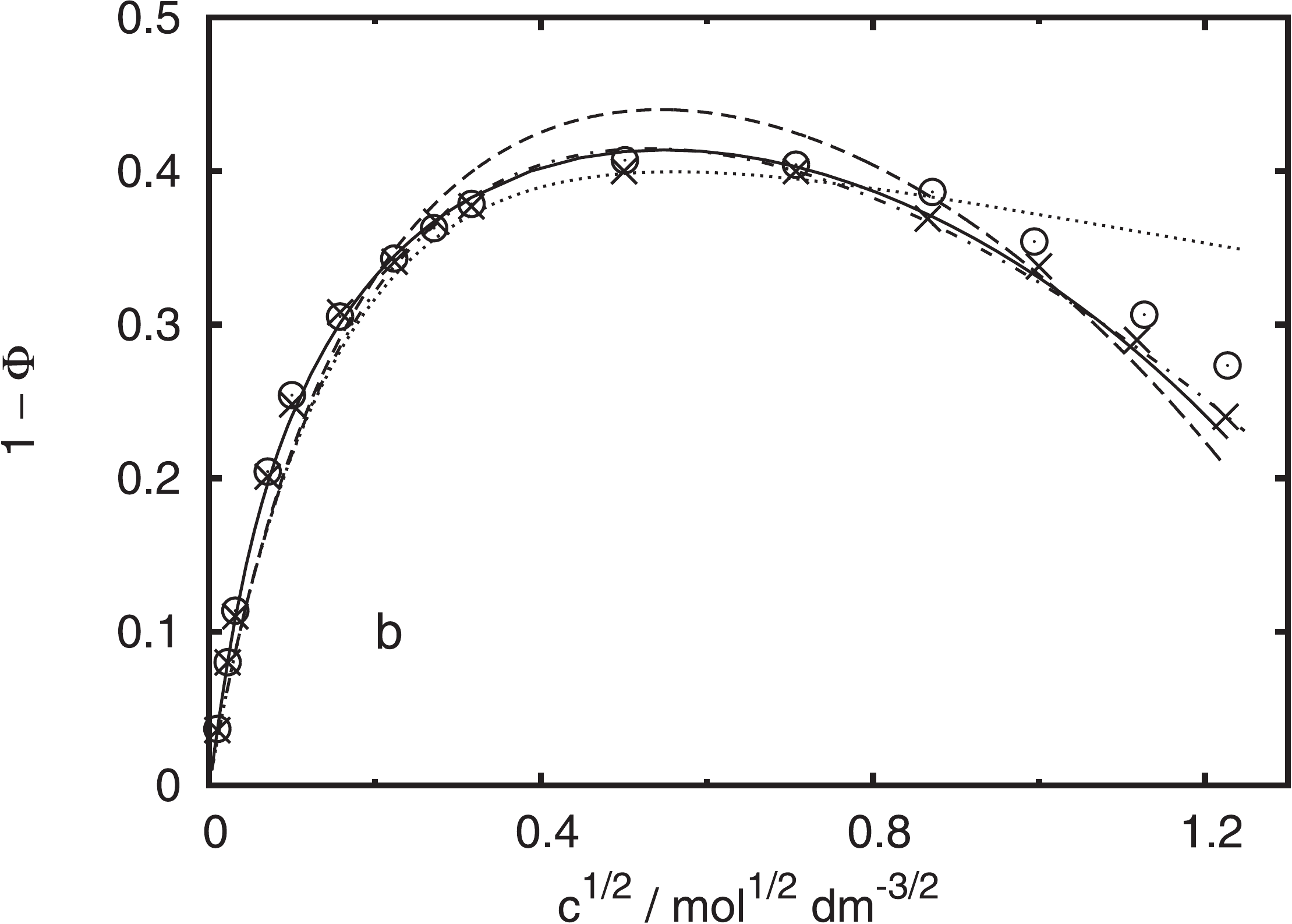}
\caption{The same as in figure~\ref{fig:11sim}, but for the
$+2$:$-2$ model electrolyte.} \label{fig:22sim}
\end{figure}

Thermodynamic properties, e.g. excess internal energies, osmotic
coefficients, and activity coefficients, are for three examples:
(a) $z_{+}= - z_{-} = 1$, $a_{+}=a_{-} = 4.25$~\AA; (b) $z_{+}= -
z_{-} = 1$, $a_{+}= 5.43$~\AA, $a_{-} = 3.62$~\AA; and (c) $z_{+}=
- z_{-} = 2$, $a_{+}=a_{-} = 4.25$~\AA, as a function of the
electrolyte molar concentration, presented in
tables~\ref{tab:hresult2}--\ref{tab:hresultx22s}, and shown in
figures~\ref{fig:11sim}--\ref{fig:22sim}. The molar concentration
of charge symmetric electrolytes, $c=c_{+}=c_{-}$ (mol dm$^{-3}$)
is related to the number densities in the following way:
$\rho_{i}$= $c_i$ $N_{\mathrm{A}}$\,,  where $N_{\mathrm{A}}$ is
the Avogadro's number. Basic equations for the mean spherical
approximation, Pitzer's approach, and the Debye-H\"uckel theory,
which are not in focus of this paper, are given in appendices.

\medskip
(a) $z_{+}= - z_{-} = 1$, $a_{+}=a_{-} = 4.25$ \AA.

\smallskip

The excess internal energy, $\beta E^\mathrm{ex} /N$, chemical
potential, $\ln \gamma_{\pm}$\,, and osmotic coefficient, $\Phi$,
at different concentrations of size symmetric $+1$:$-1$ electrolyte
are presented in tables~\ref{tab:hresult2}
and~\ref{tab:hresultx11s}. While all the GCMC results were
obtained using the Ewald summation technique
(table~\ref{tab:hresult2}), the comparison between the canonical
Monte Carlo results obtained using the periodic boundary
conditions with minimum image convention and results obtained
using Ewald summation are given in table~\ref{tab:hresultx11s}.
For simulations in the canonical ensemble, the Widom method was
used to calculate the mean activity coefficients. As seen from
table~\ref{tab:hresultx11s}, for $+1$:$-1$ electrolyte in the
concentration range studied here (from 0.0001~mol\,dm$^{-3}$ to
1.5~mol\,dm$^{-3}$), both ways of accounting for the finite sample
effects (minimum image, Ewald summation), yield the results, which
agree within 5\% or (most often) better. This appears to be within
the sum of numerical uncertainties of separate calculations.

Figure~\ref{fig:11sim} shows the results presented in
tables~\ref{tab:hresult2} and~\ref{tab:hresultx11s} ($\ln
\gamma_{\pm}$ and $1-\Phi$; i.e. deviations from ideality), as
well as the results obtained by the frequently used analytical
theories (DH approximation, Pitzer's approach, MSA theory~-- see
equations in the appendices). $\ln \gamma_{\pm}$ (panel (a)) and
$1-\Phi$ (panel (b)) are shown for different methods as a function
of square root of the concentration. It is clear from this figure
that the DH theory quantitatively describes the osmotic and activity
coefficients of the model electrolyte only up to
approximately $c=0.05$~mol\,dm$^{-3}$. Pitzer's continuation
extends the range of the DH approach up to 0.75~mol\,dm$^{-3}$.
By contrast, the MSA and HNC approximations are equally good in the
whole concentration range, only at concentrations higher than 1~M,
the HNC performs slightly better.

The main purpose of this work is to test the validity of the HVB
equation (\ref{belloni}). First we checked the thermodynamic
consistency of the HNC approach by comparing the mean activity
coefficients obtained by the Gibbs-Duhem equation with those
obtained directly (equation~(\ref{belloni})). The former results
in table~\ref{tab:hresultx11s} are denoted as $\gamma
_{\pm,\mathrm{GD}}$\,. One can see that the two sets of results
agree very well. It is our impression, however, that  the mean
activity coefficients calculated via the Gibbs-Duhem equation are
slightly closer to the MC data than those calculated by the HVB
equation. The latter results for $\gamma_{\pm}$ seem to be
systematically too high, but the differences are small, most
likely within the experimental uncertainties.

\medskip
(b) $z_{+}= - z_{-} = 1$, $a_{+}= 5.43$ \AA, $a_{-} = 3.62$ \AA.

\smallskip

The results for size asymmetric $+1$:$-1$ electrolyte are presented
in tables~\ref{tab:hresult11a} and~\ref{tab:hresultx11a} and
figure~\ref{fig:11asim}. Notice that in the DH theory calculations
the parameter $a$ was taken to be $a={\frac{1}{2}({a_{+} +
a_{-}}})$ since the DH theory cannot take the size asymmetry of
the ions into account. The conclusions are the same as for the
case (a): the size asymmetry considered here does not dramatically
effect the performance of these theories in the applied
concentration range.

\medskip
(c) $z_{+}= - z_{-} = 2$, $a_{+}=a_{-} = 4.25$ \AA.

\smallskip
The excess internal energy, $\beta E^\mathrm{ex} /N$, excess
chemical potential, $\ln \gamma _{\pm}$\,, and osmotic
coefficient, $\Phi$, for different concentrations of size
symmetric $+2$:$-2$ electrolyte are presented in
tables~\ref{tab:hresult22} and~\ref{tab:hresultx22s} and
figure~\ref{fig:22sim} (only $\ln \gamma_{\pm}$ and $\Phi$).
Interestingly, different treatments of the boundary conditions do
not effect the results for the excess chemical potential obtained
by Widom's method. This is not entirely true for the excess
internal energy and osmotic coefficient calculations at
concentrations above $1$~mol\,dm$^{-3}$. The discrepancies may
become quite large at concentrations $\approx 1.5$~mol\,dm$^{-3}$,
and the Ewald summation method should be used to obtain correct
results. For this reason the results of computer simulations,
using the minimum image convention at $1.5$~mol\,dm$^{-3}$, are
not included in table~\ref{tab:hresultx22s}.  At such a high
concentration of the $+2$:$-2$ electrolyte, the minimum image method
yields unphysical pair distribution functions, and consequently
should not be used for accurate simulations.

As seen from figure~\ref{fig:22sim}, the differences among
different theories become, as expected, more pronounced in the
case of $+2$:$-2$ electrolyte than for $+1$:$-1$ electrolyte examined
before. The results presented in this paper suggest, in close
agreement with many previous calculations, that the HNC
approximation accurately describes the thermodynamic properties of
the primitive model electrolytes in a wide concentration range.

\section{Conclusions}
The expression for the mean activity coefficient valid within the
HNC approximation, Hansen-Vieillefosse-Belloni equation, was for
primitive model $+1$:$-1$ and $+2$:$-2$ electrolytes tested against
the newly obtained Monte Carlo simulation data. For the sake of
completeness, the thermodynamic properties of $+1$:$-1$ and $+2$:$-2$
electrolytes, calculated by means of some other theories, often
used in describing the electrolyte solutions, are also presented.

Special attention is paid to the numerical accuracy of simulations
and the HNC calculations. Although the HVB equation
(expression~(\ref{belloni})) has been used before, to our best
knowledge, a systematic test of its accuracy has not been
performed so far. For all the examples studied here, the HVB
equation, compared to the canonical or grand canonical ensemble
Monte Carlo simulations, yields accurate results for the mean
activity coefficients and is, especially at higher concentration
and $+2$:$-2$ electrolytes, superior to other approximations
examined here. At the same time, it appears to be simple in usage
and hence represents an excellent tool in describing the excess
chemical potential of bulk electrolyte solutions. Notice, that
this information is needed~\cite{jamnik95,luksic2007} whenever the
membrane equilibria involving electrolyte solutions are studied.

\section*{Acknowledgements}

E.~G.-V., M.~L., B.~H.-L. and V.~V. appreciate the financial
support of the Slovenian Research Agency through grant P1--0201.
E.~G.-V. and B.~M.-M. acknowledge the support of the
supercomputing department DGSCA of the National University of
M\'{e}xico (UNAM) for the access to the Kan Balam cluster.
E.~G.-V. acknowledges CONACyT for the financial support through
the scholarship given and the Becas Mixtas program that allowed
him the stay in Slovenia.

\newpage
\begin{landscape}
%
\begin{table}[!t]
\caption{The reduced excess internal energy, $\beta
E^\mathrm{ex}/N$, the logarithm of the mean activity
coefficient, $\ln \gamma _{\pm}$, and the osmotic coefficient,
$\Phi$, as obtained using the GCMC, HNC
(equations~(\ref{energy}), (\ref{belloni}),
and~(\ref{eq:osmoznikoeficient1})), and MSA (appendix C) theory
for a $+1$:$-1$ model electrolyte. $a_{+} = a _{-} = 4.25$~\AA,
$\lambda_\mathrm{B} = 7.14$~\AA.} \centering
\vspace{1ex}
\renewcommand{\arraystretch}{0.55}
\begin{tabular}{cccccccccccc}
\hline\\[-6pt]
     &             \multicolumn{3}{c}{GCMC}           & &     \multicolumn{3}{c}{HNC} & &   \multicolumn{3}{c}{MSA}     \\
\cline{2-4} \cline{6-8} \cline {10-12}\\[-6pt]
$c$ / mol\,dm$^{-3}$  & $\beta E^\mathrm{ex} / N$ & $\ln \gamma _{\pm}$ & $\Phi$ & & $\beta E^\mathrm{ex} / N$ & $\ln \gamma_{\pm,\mathrm{HVB}}$ & $\Phi$ & & $\beta E^\mathrm{ex} / N$ & $\ln \gamma _{\pm}$ & $\Phi$ \\
\hline\\[-6pt]
$9.972\cdot10^{-5}$ & $-1.202\cdot10^{-2}$  &  $-9.21\cdot10^{-3}$  &  $0.996$  &&   $-1.147\cdot10^{-2}$ & $-1.14\cdot10^{-2}$ & $0.996$ & & $-1.156\cdot10^{-2}$   & $-1.15\cdot10^{-2}$ & $0.966$    \\
$4.986\cdot10^{-4}$ & $-2.605\cdot10^{-2}$  &  $-2.30\cdot10^{-2}$  &  $0.992$  &&   $-2.568\cdot10^{-2}$ & $-2.54\cdot10^{-2}$ & $0.992$ & & $-2.542\cdot10^{-2}$  & $-2.52\cdot10^{-2}$ & $0.992$ \\
$9.973\cdot10^{-4}$ & $-3.620\cdot10^{-2}$  &  $-3.32\cdot10^{-2}$  &  $0.988$  &&   $-3.599\cdot10^{-2}$ & $-3.53\cdot10^{-2}$ & $0.989$ & & $-3.551\cdot10^{-2}$    & $-3.51\cdot10^{-2}$ & $0.989$   \\
$4.982\cdot10^{-3}$ & $-7.707\cdot10^{-2}$  &  $-7.15\cdot10^{-2}$  &  $0.977$  &&   $-7.708\cdot10^{-2}$ & $-7.43\cdot10^{-2}$ & $0.977$ & & $-7.556\cdot10^{-2}$   &$-7.36\cdot10^{-2}$  & $0.977$    \\
$9.964\cdot10^{-3}$ & $-0.1053$            &  $-9.74\cdot10^{-2}$   &  $0.969$  &&   $-0.1055$ & $-0.100$          & $0.969$ & & $-0.1032$  &$-9.93\cdot10^{-2}$ & $0.970$  \\
$2.477\cdot10^{-2}$ & $-0.1559$            &  $-0.136$          &  $0.958$  &&  $-0.1562$ & $-0.145$           & $0.958$ & & $-0.1526$  & $-0.143$ & $0.959$    \\
$4.982\cdot10^{-2}$ & $-0.2066$            &  $-0.183$          &  $0.950$  &&  $-0.2070$ & $-0.185$           & $0.950$ & & $-0.2026$  & $-0.183$ & $0.950$    \\
$7.473\cdot10^{-2}$ & $-0.2413$            &  $-0.207$          &  $0.946$  &&  $-0.2416$ & $-0.210$           & $0.946$ & & $-0.2368$  & $-0.208$ & $0.947$    \\
$9.528\cdot10^{-2}$ & $-0.2636$            &  $-0.220$          &  $0.945$  &&  $-0.2641$ & $-0.224$           & $0.945$ & & $-0.2591$  & $-0.222$ & $0.946$    \\
$0.2490$           & $-0.3656$            &  $-0.262$           &  $0.954$  &&  $-0.3662$ & $-0.265$           & $0.955$ & & $-0.3610$  & $0.262$  & $0.955$    \\
$0.4980$           & $-0.4511$            &  $-0.246$           &  $0.991$  &&  $-0.4521$ & $-0.250$           & $0.991$ & & $-0.4465$  & $-0.245$ & $0.993$    \\
$0.7480$           & $-0.5062$            &  $-0.196$           &  $1.04$   &&  $-0.5070$ & $-0.197$           & $1.04$  & & $-0.5004$  & $-0.190$ & $1.04$ \\
$0.9976$           & $-0.5472$            &  $-0.125$           &  $1.09$   &&  $-0.5477$ & $-0.126$           & $1.09$  & & $-0.5399$  & $-0.116$ & $1.10$ \\
$1.247$            & $-0.5800$            &  $-4.00\cdot10^{-2}$    &  $1.15$   &&   $-0.5805$ & $-4.03\cdot10^{-2}$ & $1.15$  & & $-0.5711$  & $-2.75\cdot10^{-2}$ & $1.12$    \\
$1.498$            & $-0.6078$            &  $5.44\cdot10^{-2}$ &  $1.21$   &&  $-0.6084$ & $5.84\cdot10^{-2}$  & $1.22$  & & $-0.5970$  & $7.25\cdot10^{-2}$  & $1.22$ \\
\hline
\end{tabular}
\label{tab:hresult2}
\end{table}
%
%
%
%
\begin{table}[!b]
\caption{The reduced excess internal energy, $\beta
E^\mathrm{ex}/N$, the logarithm of the mean activity
coefficient, $\ln \gamma_{\pm}$, and the osmotic coefficient,
$\Phi$, as obtained using the canonical ensemble Monte Carlo method (minimum image,
Ewald summation), and HNC theory (equations~(\ref{gd})
and~(\ref{belloni}) for $\ln \gamma_{\pm}$) for a $+1$:$-1$ model
electrolyte. $a_{+} = a _{-} = 4.25$~\AA, $\lambda_\mathrm{B} =
7.14$~\AA.} \centering
\vspace{1ex}
\renewcommand{\arraystretch}{0.55}
\begin{tabular}{ccccccccccccc}
\hline\\[-6pt]
     &             \multicolumn{7}{c}{MC}           &      \multicolumn{5}{c}{HNC}    \\
\cline{2-8} \cline{10-13} \\[-6pt]
     &             \multicolumn{3}{c}{Minimum image}  & &     \multicolumn{3}{c}{Ewald summation} &    \\
\cline{2-4} \cline{6-8}   \\[-6pt]
$c$ / mol\,dm$^{-3}$  & $\beta E^\mathrm{ex} / N$ & ln $\gamma _{\pm}$ & $\Phi$ & &  $\beta E^\mathrm{ex} / N$ & ln $\gamma _{\pm}$ & $\Phi$  & &  $\beta E^\mathrm{ex} / N$ & $\ln \gamma _{\pm,\mathrm{HVB}}$  & $\ln \gamma_{\pm,\mathrm{GD}}$ & $\Phi$ \\
\hline\\[-6pt]
$0.0001$ &     $-0.01182$   &   $-0.0120$   &   $0.996$        & &         $-0.0119$  &  $-0.0115$  &  $0.996$      & &        $-0.01148$ & $-0.0114$ & $-$ & $0.9962$\\
$0.0005$ &     $-0.02611$   &   $-0.0258$   &   $0.991$         & &         $-0.0262$  &  $-0.0244$  & $0.992$      & &        $-0.02572$ & $-0.0254$ & $-$ & $0.9917$\\
$0.0010$ &     $-0.03651$   &   $-0.0358$   &   $0.988$         & &         $-0.0363$  &  $-0.0340$  &  $0.988$      & &        $-0.03603$ & $-0.0354$ & $-$ & $0.9885$\\
$0.0050$ &     $-0.07771$   &   $-0.0750$   &   $0.976$         & &         $-0.0773$  &  $-0.0719$  &  $0.977$      & &        $-0.07721$ & $-0.0744$ & $-0.0748$ & $0.9766$ \\
$0.0100$ &     $-0.1062$    &   $-0.101$    &   $0.969$         & &         $-0.1057$  &  $-0.0972$  &  $0.969$      & &        $-0.1057$  & $-0.1005$ & $-0.1010$ & $0.9692$ \\
$0.0250$ &     $-0.1574$    &   $-0.146$    &   $0.957$         & &         $-0.1568$  &  $-0.141$   &  $0.958$      & &        $-0.1568$  & $-0.1450$ & $-0.1458$  & $0.9580$ \\
$0.0500$ &     $-0.2079$    &   $-0.186$    &   $0.949$         & &         $-0.2071$  &  $-0.181$   &  $0.950$      & &        $-0.2073$  & $-0.1852$ & $-0.1860$  & $0.9498$ \\
$0.0750$ &     $-0.2424$    &   $-0.211$    &   $0.945$         & &         $-0.2416$  &  $-0.205$   &  $0.946$      & &        $-0.2419$  & $-0.2098$ & $-0.2106$  & $0.9464$ \\
$0.1000$ &     $-0.2694$    &   $-0.228$    &   $0.943$         & &         $-0.2685$  &  $-0.222$   &  $0.945$      & &        $-0.2687$  & $-0.2267$ & $-0.2275$  & $0.9451$ \\
$0.2500$ &     $-0.3673$    &   $-0.266$    &   $0.953$         & &         $-0.3664$  &  $-0.260$   &  $0.955$      & &        $-0.3667$  & $-0.2653$ & $-0.2662$  & $0.9547$ \\
$0.5000$ &     $-0.4533$    &   $-0.250$    &   $0.989$         & &         $-0.4519$  &  $-0.245$   &  $0.991$       & &        $-0.4526$  & $-0.2492$ & $-0.2500$ & $0.9918$ \\
$0.7500$ &     $-0.5081$    &   $-0.198$    &   $1.04$         & &         $-0.5067$  &  $-0.193$   &  $1.04$       & &        $-0.5073$ & $-0.1970$ & $-0.1976$    & $1.039$ \\
$1.0000$ &     $-0.5491$    &   $-0.127$    &   $1.09$         & &         $-0.5475$  &  $-0.123$   &  $1.09$       & &        $-0.5481$  & $-0.1252$ & $-0.1257$   & $1.093$ \\
$1.2500$ &     $-0.5821$    &   $-0.042$    &   $1.15$         & &         $-0.5802$  &  $-0.039$   &  $1.15$       & &        $-0.5809$ & $-0.0393$ & $-0.0393$ & $1.152$ \\
$1.5000$ &     $-0.6099$    &   $0.0533$    &   $1.21$         & &         $-0.6079$  &   $0.056$   &  $1.21$       & &        $-0.6084$  & $ 0.0584$ & $0.0588$ & $1.216$ \\
\hline
\end{tabular}
\label{tab:hresultx11s}
\end{table}
\end{landscape}

\begin{landscape}

\begin{table}[!t]
\caption{The same as in table~\ref{tab:hresult2}, but for
$a_{+} = 5.43$~\AA, $a _{-} = 3.62$~\AA.} \centering

\vspace{1ex}
\renewcommand{\arraystretch}{0.55}
\begin{tabular}{cccccccccccc}
\hline\\[-6pt]
     &             \multicolumn{3}{c}{GCMC}           & &     \multicolumn{3}{c}{HNC} & &   \multicolumn{3}{c}{MSA}     \\
\cline{2-4} \cline{6-8} \cline {10-12}\\[-6pt]
$c$ / mol\,dm$^{-3}$  & $\beta E^\mathrm{ex} / N$ & $\ln \gamma _{\pm}$ & $\Phi$ & & $\beta E^\mathrm{ex} / N$ & $\ln \gamma_{\pm,\mathrm{HVB}}$ & $\Phi$ & & $\beta E^\mathrm{ex} / N$ & $\ln \gamma _{\pm}$ & $\Phi$ \\
\hline\\[-6pt]
$9.984\cdot10^{-5}$ & $-1.185\cdot10^{-2}$  &  $-1.04\cdot10^{-2}$  &  $0.996$  &&   $-1.146\cdot10^{-2}$ & $-1.14\cdot10^{-2}$ & $0.996$ & & $-1.156\cdot10^{-2}$   & $-1.15\cdot10^{-2}$ & $0.966$    \\
$4.986\cdot10^{-4}$ & $-2.571\cdot10^{-2}$  &  $-2.29\cdot10^{-2}$  &  $0.992$  &&   $-2.562\cdot10^{-2}$ & $-2.52\cdot10^{-2}$ & $0.992$ & & $-2.538\cdot10^{-2}$  & $-2.51\cdot10^{-2}$ & $0.992$ \\
$9.977\cdot10^{-4}$ & $-3.600\cdot10^{-2}$  &  $-3.33\cdot10^{-2}$  &  $0.989$  &&   $-3.586\cdot10^{-2}$ & $-3.52\cdot10^{-2}$ & $0.989$ & & $-3.543\cdot10^{-2}$    & $-3.50\cdot10^{-2}$ & $0.989$   \\
$4.987\cdot10^{-3}$ & $-7.647\cdot10^{-2}$  &  $-7.12\cdot10^{-2}$  &  $0.977$  &&   $-7.655\cdot10^{-2}$ & $-7.34\cdot10^{-2}$ & $0.977$ & & $-7.522\cdot10^{-2}$   &$-7.28\cdot10^{-2}$  & $0.977$    \\
$9.973\cdot10^{-3}$ & $-0.1044$            &  $-9.66\cdot10^{-2}$   &  $0.970$  &&   $-0.1045$ & $-9.06\cdot10^{-2}$           & $0.970$ & & $-0.1026$  &$-9.77\cdot10^{-2}$ & $0.970$  \\
$2.491\cdot10^{-2}$ & $-0.1544$            &  $-0.138$          &  $0.960$  &&  $-0.1545$ & $-0.141$                     & $0.960$ & & $-0.1517$  & $-0.139$ & $0.960$  \\
$4.985\cdot10^{-2}$ & $-0.2036$            &  $-0.175$          &  $0.954$  &&  $-0.2038$ & $-0.177$                     & $0.954$ & & $-0.2004$  & $-0.176$ & $0.954$  \\
$7.471\cdot10^{-2}$ & $-0.2370$            &  $-0.195$          &  $0.952$  &&  $-0.2374$ & $-0.198$                     & $0.952$ & & $-0.2338$  & $-0.197$ & $0.952$  \\
$9.972\cdot10^{-2}$ & $-0.2632$            &  $-0.210$          &  $0.952$  &&  $-0.2590$ & $-0.210$                     & $0.952$ & & $-0.2599$  & $-0.210$ & $0.952$  \\
$0.2490$           & $-0.3575$            &  $-0.228$           &  $0.972$  &&  $-0.3582$ & $-0.231$                     & $0.971$ & & $-0.3550$  & $-0.229$ & $0.972$   \\
$0.4983$           & $-0.4409$            &  $-0.179$           &  $1.03$   &&  $-0.4414$ & $-0.181$                     & $1.03$  & & $-0.4381$   & $-0.177$ & $1.03$  \\
$0.7481$           & $-0.4943$            &  $-9.05\cdot10^{-2}$    &  $1.09$   &&   $-0.4945$ & $-9.16\cdot10^{-2}$             & $1.09$  & & $-0.4904$  & $-8.54\cdot10^{-2}$ & $1.10$    \\
$0.9984$           & $-0.5343$            &  $1.98\cdot10^{-2}$ &  $1.16$   &&  $-0.5342$ & $2.19\cdot10^{-2}$            & $1.17$  & & $-0.5287$  & $3.01\cdot10^{-2}$ & $1.17$    \\
$1.248$            & $-0.5665$            &  $0.150$            &  $1.25$   &&  $-0.5662$ & $0.154$                      & $1.25$  & & $-0.5589$  & $0.163$ & $1.26$    \\
$1.498$            & $-0.5936$            &  $0.295$            &  $1.34$   &&  $-0.5932$ & $0.306$                      & $1.35$  & & $-0.5839$  & $0.313$  & $1.35$   \\

\hline
\end{tabular}
\label{tab:hresult11a}
\end{table}


\begin{table}[!b]
\caption{The same as in table~\ref{tab:hresultx11s}, but for
$a_{+} = 5.43$~\AA, $a_{-} = 3.62$~\AA\ ($\ln
\gamma_{\pm,\mathrm{GD}}$ is not given).} \centering

\bigskip
\renewcommand{\arraystretch}{0.55}
\begin{tabular}{cccccccccccc}
\hline\\[-6pt]
     &             \multicolumn{7}{c}{MC}           &      \multicolumn{4}{c}{HNC}    \\
\cline{2-8} \cline{10-12} \\[-6pt]
     &             \multicolumn{3}{c}{Minimum image}  & &     \multicolumn{3}{c}{Ewald summation} &    \\
\cline{2-4} \cline{6-8} \\[-6pt]

$c$ / mol\,dm$^{-3}$  & $\beta E^\mathrm{ex} / N$ & ln $\gamma _{\pm}$ & $\Phi$ & &  $\beta E^\mathrm{ex} / N$ & ln $\gamma _{\pm}$ & $\Phi$  & &  $\beta E^\mathrm{ex} / N$ & $\ln \gamma _{\pm,\mathrm{HVB}}$  & $\Phi$ \\
\hline\\[-6pt]
$0.0001$ &     $-0.01181$   &   $-0.0120$   &   $0.996$        & &         $-0.0120$  &  $-0.0115$  &  $0.996$      & &        $-0.01147$ & $-0.0114$ & $0.9962$\\
$0.0005$ &     $-0.02611$   &   $-0.0257$   &   $0.991$        & &         $-0.0259$  &  $-0.0245$  &  $0.992$      & &        $-0.02565$ & $-0.0253$ & $0.9917$\\
$0.0010$ &     $-0.03641$   &   $-0.0356$   &   $0.988$        & &         $-0.0363$  &  $-0.0338$  &  $0.989$     & &        $-0.03590$ & $-0.0352$ & $0.9886$\\
$0.0050$ &     $-0.07711$   &   $-0.0740$   &   $0.977$       & &         $-0.0767$   &  $-0.0708$  &  $0.977$      & &        $-0.07664$ & $-0.0735$ & $0.9771$\\
$0.0100$ &     $-0.1052$    &   $-0.0993$   &   $0.969$        & &         $-0.1048$  &  $-0.0957$  &  $0.970$      & &        $-0.1047$  & $-0.0987$ & $0.9701$\\
$0.0250$ &     $-0.1553$    &   $-0.142$    &   $0.959$        & &         $-0.1547$  &  $-0.137$  &  $0.960$      & &        $-0.1548$  & $-0.1408$ & $0.9600$\\
$0.0500$ &     $-0.2047$    &   $-0.178$    &   $0.952$        & &         $-0.2039$  &  $-0.173$   &  $0.954$      & &        $-0.2040$  & $-0.1773$ &  $0.9536$\\
$0.0750$ &     $-0.2383$    &   $-0.199$    &   $0.950$        & &         $-0.2376$  &  $-0.193$   &  $0.952$     & &        $-0.2377$  & $-0.1984$ &  $0.9518$\\
$0.1000$ &     $-0.2644$    &   $-0.213$    &   $0.951$        & &         $-0.2637$  &  $-0.207$  &  $0.952$      & &        $-0.2638$  & $-0.2119$ &  $0.9521$\\
$0.2500$ &     $-0.3594$    &   $-0.232$    &   $0.969$        & &         $-0.3582$  &  $-0.225$   &  $0.97$       & &        $-0.3586$  & $-0.2305$ &  $0.9714$\\
$0.5000$ &     $-0.4428$    &   $-0.182$    &   $1.02$        & &         $-0.4413$  &  $-0.176$   &  $1.03$       & &        $-0.4418$  & $-0.1804$ &  $1.026$\\
$0.7500$ &     $-0.4961$    &   $-0.0931$   &   $1.09$        & &         $-0.4944$  &  $-0.0880$  &  $1.09$       & &        $-0.4948$  & $-0.0908$ &  $1.093$\\
$1.0000$ &     $-0.5362$    &   $0.0182$    &   $1.16$        & &         $-0.5343$  &  $-0.023$   &  $1.17$       & &        $-0.5345$  &  $0.0227$ &  $1.169$\\
$1.2500$  &    $-0.5685$    &   $0.148$        &    $1.25$        & &         $-0.5664$  &  $ 0.151$   &  $1.25$       & &        $-0.5665$  &  $0.1555$ &  $1.256$\\
$1.5000$ &    $-0.5959 $    &   $0.293$        &    $1.34$        & &         $-0.5936$  &  $0.297$   &  $1.34$       & &        $-0.5935$  &  $0.3064$ &  $1.352$\\
\hline
\end{tabular}
\label{tab:hresultx11a}
\end{table}

\end{landscape}

\newpage

\begin{landscape}

\begin{table}[!t]
\caption{The same as in table~\ref{tab:hresult2}, but for the
$+2$:$-2$ model electrolyte.} \centering

\vspace{1ex}
\renewcommand{\arraystretch}{0.55}
\begin{tabular}{cccccccccccc}
\hline\\[-6pt]
     &             \multicolumn{3}{c}{GCMC}           & &     \multicolumn{3}{c}{HNC} & &   \multicolumn{3}{c}{MSA}     \\
\cline{2-4} \cline{6-8} \cline {10-12}\\[-6pt]
$c$ / mol\,dm$^{-3}$  & $\beta E^\mathrm{ex} / N$ & $\ln \gamma_{\pm}$ & $\Phi$ & & $\beta E^\mathrm{ex} / N$ & $\ln \gamma_{\pm,\mathrm{HVB}}$ & $\Phi$ & & $\beta E^\mathrm{ex} / N$ & $\ln \gamma _{\pm}$ & $\Phi$ \\
\hline\\[-6pt]
$9.980\cdot10^{-5}$ & $-0.1266$  &  $-0.102$    &  $0.963$  &&  $-0.1264$ & $-0.104$  & $0.964$  & & $-9.126\cdot10^{-2}$   & $-9.12\cdot10^{-2}$ & $0.970$ \\
$5.000\cdot10^{-4}$ & $-0.3220$  &  $-0.238$    &  $0.920$  &&  $-0.3095$ & $-0.237$  & $0.920$  & & $-0.1978$  & $-0.198$ & $0.936$     \\
$9.968\cdot10^{-4}$ & $-0.4572$  &  $-0.333$    &  $0.886$  &&  $-0.4339$ & $-0.331$  & $0.891$  & & $-0.2728$  & $-0.272$ & $0.913$     \\
$4.982\cdot10^{-3}$ & $-0.9302$  &  $-0.678$    &  $0.796$  &&  $-0.8366$ & $-0.660$  & $0.804$  & & $-0.5573$  & $-0.555$ & $0.831$     \\
$9.963\cdot10^{-3}$ & $-1.176$   &  $-0.879$    &  $0.746$  &&  $-1.056$  & $-0.855$  & $0.759$  & & $-0.7417$  & $-0.738$ & $0.782$     \\
$2.493\cdot10^{-2}$ & $-1.516$   &  $-1.19$ &  $0.695$  &&  $-1.384$  & $-1.16$   & $0.699$  & & $-1.054$   & $-1.04$  & $0.709$    \\
$4.979\cdot10^{-2}$ & $-1.780$   &  $-1.45$ &  $0.657$  &&  $-1.658$  & $-1.43$   & $0.655$  & & $-1.342$   & $-1.32$  & $0.652$    \\
$7.345\cdot10^{-2}$ & $-1.928$   &  $-1.60$ &  $0.637$  &&  $-1.821$  & $-1.59$   & $0.633$  & & $-1.522$   & $-1.49$  & $0.622$    \\
$0.1000$           & $-2.049$   &  $-1.73$  &  $0.621$  &&  $-1.955$  & $-1.72$   & $0.617$  & & $-1.674$   & $-1.63$  & $0.601$    \\
$0.2510$           & $-2.423$   &  $-2.11$  &  $0.593$  &&  $-2.374$  & $-2.12$   & $0.587$  & & $-2.163$   & $-2.06$  & $0.561$    \\
$0.5001$           & $-2.725$   &  $-2.38$  &  $0.596$  &&  $-2.708$  & $-2.39$   & $0.597$  & & $-2.553$   & $-2.35$  & $0.575$    \\
$0.7594$           & $-2.921$   &  $-2.52$  &  $0.614$  &&  $-2.918$  & $-2.52$   & $0.630$  & & $-2.793$   & $-2.48$  & $0.617$    \\
$0.9885$           & $-3.052$   &  $-2.57$  &  $0.646$  &&  $-3.054$  & $-2.58$   & $0.667$  & & $-2.945$   & $-2.53$  & $0.665$    \\
$1.270$            & $-3.183$   &  $-2.60$  &  $0.694$  &&  $-3.187$  & $-2.60$   & $0.722$  & & $-3.089$   & $-2.53$  & $0.732$    \\
$1.506$            & $-3.292$   &  $-2.60$  &  $0.727$  &&  $-3.280$  & $-2.59$   & $0.774$  & & $-3.186$   & $-2.51$  & $0.794$    \\

\hline
\end{tabular}
\label{tab:hresult22}
\end{table}



\begin{table}[ht]
\caption{The same as in table~\ref{tab:hresultx11s}, but for
the $+2$:$-2$ model electrolyte ($\ln \gamma_{\pm,\mathrm{GD}}$
is not given).} \centering

\vspace{1ex}
\renewcommand{\arraystretch}{0.55}
\begin{tabular}{cccccccccccc}
\hline\\[-6pt]
     &             \multicolumn{7}{c}{MC}           &      \multicolumn{4}{c}{HNC}    \\
\cline{2-8} \cline{10-12} \\[-6pt]
     &             \multicolumn{3}{c}{Minimum image}  & &     \multicolumn{3}{c}{Ewald summation} &    \\
\cline{2-4} \cline{6-8} \\[-6pt]

$c$ / mol\,dm$^{-3}$  & $\beta E^\mathrm{ex} / N$ & ln $\gamma _{\pm}$ & $\Phi$ & &  $\beta E^\mathrm{ex} / N$ & ln $\gamma _{\pm}$ & $\Phi$  & &  $\beta E^\mathrm{ex} / N$ & $\ln \gamma _{\pm,\mathrm{HVB}}$  & $\Phi$ \\
\hline\\[-6pt]
$0.0001$ &     $-0.1286$      & $-0.104$    &   $0.962$        & &        $-0.128$  &  $-0.101$  &  $0.964$      & &        $-0.1266$ & $-0.1037$ & $0.9643$\\
$0.0005$ &     $-0.3201$      & $-0.238$    &   $0.914$        & &        $-0.321$  &  $-0.233$  &  $0.920$      & &        $-0.3095$ & $-0.2373$ & $0.9198$\\
$0.0010$ &     $-0.4608$      & $-0.336$    &   $0.883$        & &        $-0.460$  &  $-0.330$  &  $0.890$      & &        $-0.4345$ & $-0.3313$ & $0.8910$\\
$0.0050$ &     $-0.930$   & $-0.681$    &   $0.790$         & &        $-0.933$  &  $-0.675$  &  $0.799$      & &        $-0.8377$ & $-0.661$2 & $0.8033$\\
$0.0100$ &     $-1.176$          &  $-0.883$    &   $0.745$         & &        $-1.181$  &  $-0.882$   &  $0.752$      & &        $-1.057$  & $-0.8566$ & $0.7592$\\
$0.0250$ &     $-1.517$          &  $-1.19$ &   $0.691$         & &        $-1.521$  &  $-1.19$   &  $0.692$      & &        $-1.385$  & $-1.165$  & $0.6989$\\
$0.0500$ &     $-1.781$          &  $-1.45$ &   $0.656$         & &        $-1.784$  &  $-1.44$   &  $0.659$      & &        $-1.660$  & $-1.432$  & $0.6551$\\
$0.0750$ &     $-1.937$          &  $-1.62$ &   $0.635$         & &        $-1.940$  &  $-1.61$  &  $0.633$      & &        $-1.830$  & $-1.600$  & $0.6318$\\
$0.1000$ &     $-2.050$         &   $-1.73$ &   $0.623$         & &        $-2.051$  &  $-1.74$   &  $0.623$      & &        $-1.955$  & $-1.722$  & $0.6171$\\
$0.2500$ &     $-2.423$   & $-2.11$ &   $0.597$         & &        $-2.425$  &  $-2.12$   &  $0.600$      & &        $-2.372$  & $-2.119$  & $0.5873$\\
$0.5000$ &     $-2.730$   & $-2.38$ &   $0.601$          & &        $-2.727$  &  $-2.41$   &  $0.602$       & &        $-2.708$  & $-2.393$  & $0.5972$\\
$0.7500$ &     $-2.924$   & $-2.50$ &   $0.628$         & &        $-2.919$  &  $-2.60$   &  $0.631$      & &        $-2.912$  & $-2.519$  & $0.6283$\\
$1.0000$ &     $-3.071$   & $-2.58$ &   $0.663$         & &        $-3.060$  &  $-2.64$   &  $0.662$      & &        $-3.061$  & $-2.579$  & $0.6695$\\
$1.2500$ &     $-3.190$   & $-2.60$ &   $0.699$         & &        $-3.179$  &  $-2.71$   &  $0.710$       & &        $-3.179$  & $-2.599$  & $0.7179$\\
$1.5000$ &     $-$            & $-$         &   $-$                & &        $-3.278$  &  $-2.65$   &  $0.760$       & &        $-3.277$  & $-2.591$  & $0.7727$\\
\hline
\end{tabular}
\label{tab:hresultx22s}
\end{table}

\end{landscape}

\appendix

\section{Debye-H\"{u}ckel theory}
\renewcommand{\theequation}{A\arabic{equation}}

In the framework of the Debye-H\"{u}ckel theory, the expression
for the reduced excess internal energy reads
\begin{equation}
\frac{\beta E_{\mathrm{DH}}^{\mathrm{ex}}}{N} =
- \frac{\lambda_{\mathrm{B}} |z_{+} z_{-}|}{2}  \frac{\kappa}{1+\kappa a}\,,
\label{dhe}
\end{equation}
where $z_+$ and $z_-$ are valencies of cations and anions,
respectively, $\kappa^{-1}$ is the so-called Debye screening
length defined as $\kappa^2 = 4 \pi \lambda_{\mathrm{B}} \sum
\rho_i z_i^2$ ($\lambda_{\mathrm{B}}$ being the Bjerrum length
and $\rho _i$ the number density of the ionic species $i$), and
$a$ is the distance of closest approach of two ions (assumed to
be same for all pairs).

The osmotic coefficient is given by
\begin{equation}
\Phi_{\mathrm{DH}} = 1 - \frac{\lambda_\mathrm{B} |z_{+} z_{-}| \kappa}{6} \cdot
\sigma (\kappa a),
\label{dhphi}
\end{equation}
where $\sigma (\kappa a)$ is a function defined
as~\cite{robinson_stokes,pitzer77}
\begin{equation}
\sigma (x) = {\frac {3}{x^3}} \left(1 + x - {\frac{1}{1+x}} - 2
\ln (1+x)\right). \label{sigma}
\end{equation}

The mean activity coefficient is calculated via
\begin{equation}
\ln\gamma_{\pm,\mathrm{DH}}= - \frac{2.303A |z_+ z_- |\sqrt{I}}{1 + B a \sqrt{I}}\,,
\label{dha}
\end{equation}
where $I$ is the ionic strength of the solution ($I = 0.5 \sum_i
c_i z_i^2$\,, $c_i$ being the molar concentration of species $i$),
and $A$, $B$ are constants containing the absolute temperature and
dielectric constant of the solvent. For aqueous solutions at
25~$^{\circ}$C, $A=0.511$~dm$^{3/2}$\,mol$^{-1/2}$, and
$B=0.329\cdot 10^{-8}$~cm$^{-1}$\,dm$^{3/2}$\,mol$^{-1/2}$.

At extreme dilution, the term $B a \sqrt{I}$ in the denominator of
equation~(\ref{dha}) becomes negligible compared to unity and the
equation~(\ref{dha}) yields the known Debye-H\"{u}ckel's limiting
law (DHLL)
\begin{equation}
\ln\gamma_{\pm,\mathrm{DHLL}}= - 2.303 A |z_+z_-|\sqrt{I} \,.
\label{dhll}
\end{equation}
The expression for the
osmotic coefficient reduces in this case to~\cite{lewis_randall}
\begin{equation}
\Phi_{\mathrm{DHLL}} = 1 - \frac{2.303}{3} A | z_+z_-|\sqrt{I} \,.
\label{dhphill}
\end{equation}

\section{Pitzer's approach}
\renewcommand{\theequation}{B\arabic{equation}}

The osmotic coefficient is obtained via the virial route, which
yields~\cite{pitzer77}
\begin{equation}
\Phi_{\mathrm{P}} = 1 - \frac{\lambda_{\mathrm{B}} |z_+ z_-|}{6}
\frac{\kappa}{1+\kappa a} + \frac{2}{3} \pi \rho a^{3} +
\frac{\lambda_{\mathrm{B}} |z_+ z_-|}{12 a} \left[\frac{\kappa
a}{1+\kappa a}\right]^{2}. \label{pitzerphi}
\end{equation}
In this expression $\rho = \rho_{+} + \rho_{-}$ is the total
number density of ionic species (for significance of other
quantities see appendix~A). An important improvement above the
Debye-H\"uckel theory is the last term in
equation~(\ref{pitzerphi}). The term provides the correction to
the Debye-H\"{u}ckel theory at intermediate concentrations and it
causes the osmotic coefficient to increase at higher
concentrations, exactly as observed experimentally.

The equation for the mean activity coefficient is~\cite{pitzer77}
\begin{equation}
\ln \gamma_{\pm, \mathrm{P}} = - \frac{\lambda_{\mathrm{B}} |z_+ z_-|}{6}
\frac{\kappa}{1+\kappa a} + \frac{2}{3} \pi \rho a^{3} -
\frac{|z_+ z_-| \lambda_\mathrm{B}}{6 a} \ln (1 + \kappa a) + \Phi_{\mathrm{P}} - 1,
\label{pitzera}
\end{equation}
where $\Phi _{\mathrm{P}}$ is given by the
equation~(\ref{pitzerphi}).

\section{Mean spherical approximation}
\renewcommand{\theequation}{C\arabic{equation}}

Within the MSA closure, one calculates the excess internal
energy as~\cite{blum77}
\begin{equation}
\beta E_{\mathrm{MSA}}^{\mathrm{ex}} = - \frac{\alpha^2}{4\pi}
\left[\Gamma \sum_{i} \left ( \frac{\rho_i z_i^2}{1+\Gamma a_i} \right )
+ \frac{\pi}{2\Delta}\Omega P_n^2 \right ],
\label{emsa}
\end{equation}
where $\alpha^2 = 4\pi\lambda_\mathrm{B}$\,, $\Delta = 1 -
\frac{\pi}{6}\sum_i \rho_i a_i^3$\,, and
\begin{align}
\Omega &= 1 + \frac{\pi}{2\Delta}\sum_{i} \frac{\rho_i a_i^3}{1 + \Gamma a_i}\,,\\
P_n &= \frac{1}{\Omega}\sum_{i} \frac{\rho_i a_i z_i}{1 + \Gamma a_i} \,,\\
4\Gamma^2 &= \alpha^2 \sum_i \rho_i \left [\frac{ z_i -
\frac{\pi}{2\Delta}a_i^2 P_n}{1 + \Gamma a_i} \right ]^2.
\end{align}

The electrostatic contribution to the osmotic coefficient is
given by~\cite{blum77}
\begin{equation}
\Phi^\mathrm{el} = - \frac{\Gamma}{3\pi\rho} -
\frac{\alpha^2}{8\rho} \cdot \left ( \frac{P_n}{\Delta} \right )^2,
\end{equation}
where $\rho = \sum_i \rho_i$ is the total density of the
system. The electrostatic contribution to the activity
coefficient of the species $i$ has the
form~\cite{corti87,blum77}
\begin{equation}
\ln \gamma_i^{\mathrm{el}} = \frac{\alpha^2}{4\pi}z_iM_i -
\frac{P_n a_i}{4\Delta} \left (\Gamma b_i + \frac{\pi}{12\Delta}\alpha^2 P_n a_i \right ),
\end{equation}
where
\begin{align}
b_i &= \frac{ \alpha^2 \left (z_i - \frac{\pi}{2\Delta}a_i^2 P_n \right )}
{2\Gamma (1 +  \Gamma a_i)}\,,\\
M_i &= \frac{{2\Gamma b_i}/{\alpha^2} - z_i}{a_i}\,.
\end{align}

The hard sphere contribution to the osmotic and activity
coefficients follow from the equation of state of a mixture of
hard spheres. Here we use the Mansoori-Carnahan-Starling-Leland
equation of state~\cite{mansoori71}. In this case, the osmotic
coefficient reads~\cite{mansoori71}
\begin{equation}
\Phi^\mathrm{hs} = \frac{(1 + \eta + \eta^2) - 3\eta(y_1 + y_2\eta) - \eta^3y_3}
{\Delta^3}
\label{eq:phihscs}
\end{equation}
whereas the hard sphere contribution to the activity
coefficient of the species $i$ is~\cite{ebeling83}
\begin{align}
\ln \gamma_i^\mathrm{hs} &= \left (\mu_i - 1 - \frac{2\eta_i}{\eta}y_3 \right )
 \cdot \ln \Delta \,\,  \nonumber \\
&+ \frac{\eta}{\Delta^2} \left [ 3(1 - \alpha_i) + \mu_i
+ \frac{3\eta}{2}(\alpha_i - \beta_i - \mu_i - 1) \right ] \nonumber \\
& +  \frac{\eta_i}{\Delta^3} \left \{ \eta \left [ 5y_3 -
\frac{9}{2}y_1 - 2 + \eta \left ( \frac{3}{2}y_1 - 3y_2 - 4y_3 + 1
\right ) \right ] - 2y_3 + 4 \right \}. \label{eq:lngammahscs}
\end{align}
Notations in equations~(\ref{eq:phihscs})
and~(\ref{eq:lngammahscs}) denote the following
\begin{align*}
y_1 &= \sum_{j>i} \Delta_{ij} \frac{a_i + a_j}{\sqrt{a_i a_j}} \,, \\
y_2 &= \sum_{j>i} \Delta_{ij} \sum_k \frac{\eta_k \rho_k}{\eta \rho}
\frac{\sqrt{a_i a_j}}{a_k} \,, \\
y_3 &= \left [ \sum_i \frac{\rho_i}{\rho}
\left ( \frac{\eta_i}{\eta} \right )^{2/3} \right ]^3 , \\
\Delta_{ij} &= \frac{\rho_i \rho_j}{\rho} \cdot
\frac{\sqrt{\eta_i \eta_j}}{\eta} \cdot \frac{(a_i - a_j)^2}{a_i a_j}  \,, \\
\alpha_i &= \frac{\rho}{\rho_i} \sum_k \frac{a_i + a_k}{\sqrt{a_i a_k}}
\Delta_{ik} \nonumber \,, \\
\beta_i &= \sum_{j>k} \Delta_{jk} \frac{\eta_i}{\eta}\frac{\sqrt{a_j a_k}}{a_i} +
\frac{\rho}{\rho_i} \sum_j \Delta_{ij} \sum_k \frac{\eta_k \rho_k}{\eta \rho}
\frac{\sqrt{a_i a_j}}{a_k}  \,, \\
\mu_i &= 3 \left (\frac{\eta_i}{\eta} y_3 \right )^{2/3},
\end{align*}
and as usual
\begin{equation*}
\eta_i = \frac{\pi}{6} \rho a_i^3 \,, \qquad  \eta = \sum_i \eta_i
\frac{\rho_i}{\rho} = \frac{\pi}{6} \sum_i \rho_i a_i^3 \,,
\qquad \rho = \sum_i \rho_i \,.
\end{equation*}

The osmotic and activity coefficients of the primitive model electrolyte solution
are then obtained by summing up the electrostatic and hard sphere contributions
\begin{align*}
\Phi_\mathrm{MSA} &= \Phi^\mathrm{el} + \Phi^\mathrm{hs}, \\
\ln \gamma_{i,\mathrm{MSA}} &= \ln \gamma_i^{\mathrm{el}} + \ln
\gamma_i^{\mathrm{hs}}.
\end{align*}
%
%

\ukrainianpart

\title{Примітивні моделі електролітів. Порівняння результатів гіперланцюгового наближення для коефіцієнту активності\\ з даними Монте-Карло}

\author{Е.~Гутієррез-Валладарес\refaddr{label1,label2},
М.~Лукшіч\refaddr{label2}, Б.~Міллан-Мало\refaddr{label1},
Б.~Грібар-Лі\refaddr{label2}, В.~Влахи\refaddr{label2}}

\addresses{
\addr{label1} Центр прикладної фізики та передових технологій, Національний автономний університет Мехіко, А.С. 1-1010, 76000 Керетаро, Мексика
\addr{label2} Університет Любляни, факультет хімії та хімічних технологій, Аскерчева 5, SI--1000, \\ Любляна, Словенія
}

\makeukrtitle

\begin{abstract}
\tolerance=3000%
Точність виразу для коефіцієнта середньої активності (рівняння Гансена-Віллефосса-Беллоні), справедливого у гіперланцюговому (ГЛ) наближенні, перевірялася у широкому інтервалі концентрації відносно нових даних Монте-Карло (МК) для примітивних моделей електроліту $+1$:$-1$ і $+2$:$-2$. Вираз містить ту перевагу, що надлишковий хімічний потенціал можна отримати прямо, не вдаючись до трудомісткого розрахунку Гіббса-Дюгема. Ми виявили, що результати ГЛ наближення для коефіцієнта середньої активності добре узгоджуються з проведеними для такої самої моделі числовими розрахунками. Крім того, було протестовано термодинамічну узгодженість ГЛ наближення. Виглядає так, що коефіцієнти середньої активності, пораховані за допомогою рівняння Гіббса-Дюгема, узгоджуються з даними Монте-Карло трохи краще ніж вираз Гансена-Віллефосса-Беллоні. Для цілісності розрахунку представлено також ГЛ надлишкові внутрішні енергії та осмотичні коефіцієнти. Ці результати порівнюються з розрахунками на основі інших загальновідомих теорій, які описують розчини електролітів, зокрема зі середньо-сферичним наближенням, модифікацією Пітцера теорії Дебая-Гюккеля, і граничним законом Дебая-Гюккеля.
\keywords примітивна модель електроліту, коефіцієнт середньої активності, гіперланцюгове наближення, середньо-сферичне наближення, моделювання Монте-Карло, метод Пітцера, теорія Дебая-Гюккеля
\end{abstract}


\begin{thebibliography}{70}

\bibitem{harned-owen} Harned H.S., Owen B.B., The
    Physical Chemistry of Electrolyte
    Solutions. 3rd ed., Reinhold, New York, 1958.

\bibitem{robinson-stokes} Robinson R.A., Stokes R.H.,
    Electrolyte Solutions. 2nd ed., Dover
    Publications, Mineola, New York, 2002, 432--455.

\bibitem{barthel88} Barthel J.M.C., Krienke H. -- In: Topics in Physical
    Chemistry.  Vol~5, eds. Baumg\"{a}rtel H., Franck E.U., Gr\"{u}mbein W.,
    Springer, New York, 1998.

\bibitem{Bockris1998} Bockris J.O'M., Reddy A.K.N., Modern
    Electrochemistry, vol.~1, Ionics. 2nd ed., Plenum Press, New York,
    1998.

\bibitem{loebe97} Loebe J.R., Donohue M.D., AIChE J., 1997, \textbf{43},
    180--195; \doi{10.1002/aic.690430121}.

\bibitem{Lamm2003} Lamm~G. -- In: Reviews in Computational Chemistry. Vol.~19, Ch.~4,
    147--365, eds.
    Lipkowitz K.B., Larter R., Cundari T.R.,  Wiley-VCH, John Wiley \& Sons, Inc., 2003.

\bibitem{fennell} Fennell C.J., Bizjak A., Vlachy~V., Dill~K.A.,
J.~Phys. Chem.~B, 2009, \textbf{113}, 6782--6791;\\
\doi{10.1021/jp809782z}.

\bibitem{holovko} Holovko M.F., Kapko V.I.,
Condens. Matter Phys., 2007, \textbf{10}, 397--406.

\bibitem{simonin98} Simonin J.P., Bernard O., Blum L., J.~Phys. Chem.~B,
    1998, \textbf{102}, 4411--4417; \doi{10.1021/jp9732423}.

\bibitem{hribar05} Hribar-Lee B., Vlachy V., J.~Mol. Liq., 2005, \textbf{118},
    163--169; \doi{10.1016/j.molliq.2004.07.033}.

\bibitem{debye27} Debye P., H\"{u}ckel E., Phys.~Z., 1927, \textbf{24},
    185--206.

\bibitem{valleau80} Valleau J.P., Cohen K., Card D.N., J.~Chem. Phys.,
    1980, \textbf{72}, 5942--5954; \doi{10.1063/1.439093}.

\bibitem{friedman62} Friedman H.L., Ionic Solution Theory. Wiley Interscience,
    New York, 1962.

\bibitem{Outhwaite1975}  Singer K. and Outhwaite C.W., Statistical Mechanics,
    1975, {\bf 2}, 188--255; \\ \doi{10.1039/9781847556936-00188}.

\bibitem{Fixman} Fixman M., J.~Chem. Phys., 1979, \textbf{70}, 4995--5001;
\doi{10.1063/1.437340}.

\bibitem{abbas07} Abbas Z., Ahlberg E., Nordholm S., Fluid Phase Equil.,
    2007, \textbf{260}, 233--247; \\ \doi{10.1016/j.fluid.2007.07.026}.

\bibitem{pitzer77} Pitzer K.S., Acc. Chem. Res., 1977 \textbf{10}, 371--377;
\doi{10.1021/ar50118a004}.

\bibitem{Martinez1990} Martinez M.M., Bhuiyan L.B., Outhwaite C.W.,
     J.~Chem. Soc., Faraday Trans., 1990, \textbf{86}, 3383--3390; \\
     \doi{10.1039/ft9908603383}.

\bibitem{Outhwaite1991} Outhwaite C.W., Molero M., Bhuiyan L.B.,
    J.~Chem. Soc., Faraday Trans. 1991, \textbf{87}, 3227--3230;\\ \doi{10.1039/ft9918703227}.

\bibitem{Outhwaite1993} Outhwaite C.W., Molero M., Bhuiyan L.B.,
    J.~Chem. Soc., Faraday Trans., 1993,
    \textbf{89}, 1315--1320;\\ \doi{10.1039/ft9938901315}.

\bibitem{Outhwaite2004} Outhwaite C.W., Condens. Matter Phys., 2004,
    \textbf{7}, 719--733.

\bibitem{Molero1992} Molero M., Outhwaite C.W., Bhuiyan L.B., J.~Chem. Soc.,
    Faraday Trans., 1992,
    \textbf{88}, 1541--1547;\\ \doi{10.1039/ft9928801541}.

\bibitem{Outhwaite2010} Outhwaite C.W., Bhuiyan  L.B., Vlachy~V.,
    Hribar-Lee~B., J.~Chem. Eng. Data, 2010,  \textbf{55} , 4248-4254;\\ \doi{10.1021/je100394d}.


\bibitem{hansen06} Hansen J.-P., McDonald I.R., Theory of Simple
Liquids. Elsevier, London, 2006;\\
\doi{10.1016/B978-012370535-8/50012-4}.

\bibitem{mcquarrie73} McQuarrie D.A., Statistical Mechanics. Carper Collins,
    New York, 1973.

\bibitem{blum75} Blum L., Mol. Phys., 1975, \textbf{30}, 1529--1535;
\doi{10.1080/00268977500103051}.

\bibitem{waisman72}  Waisman E., Lebowitz J.L., J.~Chem. Phys., 1972,
    \textbf{56}, 3086--3093; \doi{10.1063/1.1677644}; ibid., 3093--3099;
    \doi{10.1063/1.1677645}.

\bibitem{sanchezcastro89} Sanchez-Castro C., Blum L., J.~Phys. Chem., 1989,
    \textbf{93}, 7478--7482; \doi{10.1021/j100358a043}.

\bibitem{ebeling83}  Ebeling W., Scherwinsky K., Z.~Phys. Chem. (Leipzig),
    1983, \textbf{264}, 1--14.

\bibitem{blum77} Blum L., H{\o}ye J.S., J.~Phys. Chem., 1977, \textbf{81},
    1311--1316; \doi{10.1021/j100528a019}.

\bibitem{triolo76} Triolo R., Griera J.R., Blum L., J.~Phys. Chem., 1976,
    \textbf{80}, 1858--1861; \doi{10.1021/j100558a008}.

\bibitem{triolo78} Triolo R., Blum L., Floriano M.A., J.~Phys. Chem., 1978,
    \textbf{82}, 1368--1370; \doi{10.1021/j100501a009}.

\bibitem{corti87} Corti H.R., J.~Phys. Chem., 1987, \textbf{91}, 686--689;
\doi{10.1021/j100287a037}.

\bibitem{cartailler92} Cartailler T., Turq P., Blum L., Condamine N.,
    J.~Phys. Chem. 1992, \textbf{96}, 6766--6772;\\ \doi{10.1021/j100195a044}.

\bibitem{vilarino01}  Vilarino T., Barriada J.L., Sastre de Vicente~M.E.,
    Phys. Chem. Chem. Phys., 2001, \textbf{3}, 1053--1056;\\ \doi{10.1039/b010103f}.

\bibitem{fawcett97} Fawcett W.R., Tikanen A.C., Henderson D.J.,
    Can. J.~Chem., 1997, \textbf{75}, 1649--1655;\\ \doi{10.1139/v97-196}.

\bibitem{vilarino97} Vilarino T., Fiol S., Armesto X.L., Brandariz~I.,
    Sastre de Vicente~M.E.,
    J.~Chem. Soc., Faraday Trans., 1997, \textbf{93},  413--417; \doi{10.1039/a605917a}.

\bibitem{vilarino96} Vilarino T., Sastre de Vicente~M.E., J.~Phys. Chem.,
    1996, \textbf{100}, 16378--16384;\\ \doi{10.1021/jp9609996}.

\bibitem{vilarino99} Vilarino T., Sastre de Vicente~M.E.,
    Phys. Chem. Chem. Phys., 1999, \textbf{1}, 2453--2456;\\ \doi{10.1039/a900918c}.

\bibitem{vilarino97_jsc} Vilarino T., Sastre de Vicente~M., J.~Solut. Chem.,
    1997, \textbf{26}, 833--846; \doi{10.1007/BF02768261}.

\bibitem{Kovalenko1999} Kovalenko A., Hirata F., J.~Chem. Phys., 1999,
    \textbf{110}, 10095--10112; \doi{10.1063/1.478883}.

\bibitem{Schmeer2010} Schmeer G., Maurer A., Phys. Chem. Chem. Phys., 2010,
    \textbf{12}, 2407--2417; \doi{0.1039/B917653E}.

\bibitem{Simonin1996} Simonin J.-P., Blum L., Turq P., J.~Phys. Chem., 1996,
    \textbf{100}, 7704--7709; \doi{10.1021/jp953567o}.

\bibitem{Fawcett1996} Fawcett W.R., Tikanen A.C., J.~Phys. Chem., 1996,
    \textbf{100}, 4251--4255; \doi{10.1021/jp952379v}.

\bibitem{Vincze2010} Vincze J., Valisko M., Boda D., J.~Chem. Phys., 2010,
    \textbf{133}, 154507:1--6; \doi{10.1063/1.3489418}.

\bibitem{Blum1995} Blum L., Bernard O., J.~Stat. Phys., 1995, \textbf{79},
    569-583; \doi{10.1007/BF02184871}.

\bibitem{Bernard1996} Bernard O., Blum L., J.~Chem. Phys., 1996,
    \textbf{104}, 4746--4754; \doi{10.1063/1.471168}.

\bibitem{Rasaiah1988} Rasaiah J.C.~-- In: The Liquid State and its Electrical
    Properties. NATO ASI Series~B, {Vol. 193}, eds. Kunhardt~E.E., Christophorou~L.G., Luessen~L.H., Plenum Press, New York, 1988.

\bibitem{vlachy99} Vlachy V., Annu. Rev. Phys. Chem., 1999,
    \textbf{50}, 145--165; \doi{10.1146/annurev.physchem.50.1.145}.

\bibitem{martinez90}  Martinez M.M., Bhuiyan L.B., Outhwaite~C.W.,
    J.~Chem. Soc. Faraday Trans., 1990, \textbf{86}, 3383--3390; \\ \doi{10.1039/ft9908603383}.

\bibitem{vlachy10} Vlachy V., Hribar-Lee B., Bhuiyan L.B.,
    J.~Chem. Eng. Data, 2010, \textbf{55}, 1855--1859;\\ \doi{10.1021/je900873v}.

\bibitem{verlet62} Verlet L., Levesque D., Physica, 1962,
    \textbf{28}, 1124--1142; \doi{0.1016/0031-8914(62)90058-7}.

\bibitem{hansen76} Hansen J.-P., Vieillefosse P., Phys. Rev. Lett., 1976,
    \textbf{37}, 391--394; \doi{10.1103/PhysRevLett.37.391}.

\bibitem{belloni85} Belloni L., Chem. Phys., 1985 \textbf{99}, 43--54;
\doi{10.1016/0301-0104(85)80108-7}.

\bibitem{jamnik95} Jamnik B., Vlachy V., J.~Am. Chem. Soc., 1995,
    \textbf{117}, 8010--8016; \doi{10.1021/ja00135a020}.

\bibitem{Kjellander1989} Kjellander R., Sarman S., J.~Chem. Phys., 1989,
    \textbf{90}, 2768--2775; \doi{10.1063/1.455924}.

\bibitem{abbas09} Abbas Z., Ahlberg E., Nordholm S., J.~Phys. Chem. B, 2009,
    \textbf{113}, 5905--5916; \doi{10.1021/jp808427f}.

\bibitem{allen_tildesley} Allen M.P., Tildesley D.J., Computer
    Simulations of Liquids. Oxford University, New York, 1989.

\bibitem{widom63} Widom B., J.~Chem. Phys., 1963, \textbf{39}, 2808--2812;
\doi{10.1063/1.1734110}.

\bibitem{svensson91} Svensson B.R., Akesson T., Woodward C.E.,
    J.~Chem. Phys., 1991, \textbf{95},  2717--2726;\\ \doi{10.1063/1.460923}.

\bibitem{svensson1988} Svensson B.R., Woodward C.E., Mol. Phys., 1988,
    \textbf{64}, 247--259; \doi{10.1080/00268978800100203}.

\bibitem{valleau1980} Valleau J.P., Cohen L.K., J.~Chem. Phys., 1980,
    \textbf{72}, 5935--5941; \doi{10.1063/1.439092}.

\bibitem{jamnik1993} Jamnik B., Vlachy V., J.~Am. Chem. Soc., 1993,
    \textbf{115}, 660--666; \doi{10.1021/ja00055a040}.

\bibitem{ichiye88} Ichiye T., Haymet A.D.J., J.~Chem. Phys., 1988,
    \textbf{89}, 4315--4324; \doi{10.1063/1.454815}.

\bibitem{Rasaiah1970} Rasaiah J., Chem. Phys. Lett., 1970, \textbf{7},
    260--264; \doi{10.1016/0009-2614(70)80303-7}.

\bibitem{Stillinger1968} Stillinger F.H., Lovett~R., J.~Chem. Phys., 1968, \textbf{48},
    3858--3868; \doi{10.1063/1.1669709}.

\bibitem{robinson_stokes} Robinson R.A., Stokes R.H.,
    Electrolyte Solutions. Dover, New York, 2002, 73--86.

\bibitem{lewis_randall} Lewis G.N., Randall M.,
    Thermodynamics. McGraw Hill, New York, 1961, 332--348.

\bibitem{mansoori71} Mansoori G.A., Carnahan N.F., Starling~K.E.,
    Leland~Jr.~T.W., J.~Chem. Phys., 1971,
    \textbf{54}, 1523--1525;\\ \doi{10.1063/1.1675048}.

\bibitem{luksic2007} Luk\v{s}i\v{c}~M., Hribar-Lee~B., Vlachy~V.,
    J.~Phys. Chem.~B, 2007, \textbf{111}, 5966--5975; \doi{10.1021/jp065685p}.

\end{thebibliography}
\end{document}